\documentclass[12pt]{article}

\usepackage[utf8]{inputenc}
\usepackage[margin=1in,footskip=0.25in]{geometry}

\usepackage{amsfonts}
\usepackage{amsmath}
\usepackage{bm}
\usepackage{indentfirst}
\usepackage{graphicx}
\usepackage{float}
\usepackage{setspace}
\usepackage{textcomp}
\usepackage{makecell}
\usepackage{multirow}
\usepackage{amssymb}
\usepackage{color}
\usepackage{subcaption}
\usepackage{authblk}

\newtheorem{proposition}{Proposition}

\usepackage[pdftex, colorlinks, urlcolor=blue, linkcolor=blue, anchorcolor=blue, citecolor=blue]{hyperref}

\usepackage[round]{natbib}

\title{\Large{A Dynamical Model for Spatio-Temporal Processes Motivated by Second-Order Partial Differential Equations}}

\author[]{Yutong Zhang} 
\author[]{Xiao Liu} 
\affil[]{H. Milton Stewart School of Industrial and Systems, \\Georgia Institute of Technology}

\date{}

\begin{document}

\maketitle

\vspace{-30pt}
\begin{abstract}
An important class of spatio-temporal models is constructed by leveraging the hierarchical structure of dynamical
(or, state-space) models. This paper proposes a new statistical dynamical model for spatio-temporal processes motivated by second-order stochastic partial differential equations (SPDE). In particular, an infinite-dimensional linear state-space representation is obtained where the state transition is governed by a proposed SDE. Then, using the Galerkin's method, a finite-dimensional approximation to the infinite-dimensional SDE is obtained, yielding a dynamical model with finite states that facilitates computation and parameter estimation. The space–time covariance of the approximated dynamical model is obtained, and the error between the approximate and exact covariance matrices is quantified. Comprehensive numerical investigations, including 2D wave equation, seismic wave propagation, advection-diffusion equations and wildfire aerosol propagation processes, are performed to demonstrate the application of the proposed model. Code is available.
\end{abstract}


\singlespacing

\section{Introduction}\label{sec:Intro}

Statistical spatio-temporal models are widely used in science and engineering to describe dynamic fields such as pollutant dispersion, heat transport, fluid dynamics, etc. \citep{CressieWikle2011,Woody2022,katzfuss2024scalable,hector2024distributed,lan2025spatiotemporal}. An important class of spatio-temporal models is constructed by leveraging the hierarchical structure of dynamical (or, state-space) models; see \cite{wikle2010general, CressieWikle2011, Sigrist2015, liu2022statistical}. The state equation provides an effective integration of underlying physics, while the observation equation links the latent physical processes to observed data. 
To strengthen the process modeling and monitoring capabilities of such an important approach, 
this paper proposes a new dynamical model for spatio-temporal processes motivated by second-order stochastic partial differential equations (SPDE), and demonstrates its performance through comprehensive numerical investigations.

\vspace{4pt}
The power of dynamical spatio-temporal models comes from the dynamical model specification by a series of conditional models that lead to a complex joint space-time covariance structures which can hardly be directly specified \citep{Wikle1999, Berliner2003, CressieWikle2011}. 
Another key advantage of dynamical spatio-temporal models
arises from the natural connection to stochastic partial differential equations (SPDEs). SPDEs are a powerful framework for modeling systems where physical laws are subject to inherent randomness or fluctuations (e.g., heat transfer, wave propagation, fluid turbulence, wildfire spread, etc.). By incorporating a noise term into standard PDEs, they bridge the gap between deterministic continuum mechanics and stochastic processes. 
Because it is a common strategy to convert SPDEs into a system of stochastic differential equations (SDEs)---referred to as discretization in space---in solving SPDEs, a single SPDE governing a continuous field is transformed into a finite set of coupled equations governing discrete points or modes by approximating the spatial operators \citep{brunton2022data}. Hence, the finite set of SDEs can be naturally integrated into the dynamical model as the state transition equations that govern the system dynamics, embedding physics into statistical models. 

The approach above largely motivated a number of statistical models for physics processes. To provide a few examples, convolution-generated space-time models were described in \cite{Brown2000}. Taking the form of an integro-difference equation (IDE), a spatial field at a time is obtained from the field at the previous time step through a convolution operation with a Gaussian kernel. After each convolution operation, a white-in-time and spatially colored Gaussian noise is added to the process. Starting from some initial condition, it is shown that a non-separable stationary process can be obtained by a linear operation (i.e., a sequence of convolution operations in space and time) on spatially correlated Brownian motion through an IDE representation \citep{richardson2020spatiotemporal}. The IDE model yields a hierarchical spatio-temporal dynamical model after appropriate parameterizations; see \cite{wikle2010general}. As shown in \cite{Sigrist2015} and \cite{liu2022statistical}, the stationary process obtained from the convolution-generated model coincides with the solutions from a linear stochastic advection-diffusion equation with a Gaussian noise. In fact, constructing GP from SPDE is an effective approach for modeling physical and natural processes. From the early work of \cite{whittle1954stationary} to some more recent investigations by \cite{lindgren2011explicit, pfortner2022physics, clarotto2024spde, li2024parameter, wei2024extended, hu5149007deepomamba}, GP remains one of the most popular approaches for building surrogate models for physical and natural processes \citep{morris2012gaussian, gramacy2020surrogates, zhang2021mixed, deng2025design}. 

\vspace{4pt}
However, most of the SPDE-based statistical model consider first-order differential equations involving only first-order time derivative. In this paper, we focus on a dynamical spatio-temporal model  motivated from second-order SPDEs that involve second-order time derivative. This type of SPDE involves a range of important physical processes, including the classical wave equation, the Telegrapher's equations that model the voltage and current on an electrical transmission line over distance and time, the dynamic version of the Euler-Bernoulli beam equation, etc. In general, the second-order SPDEs  appear in systems involving inertia, acceleration, or oscillatory behavior.


In our proposed model, the second-order SPDE is firstly written as an infinite-dimensional SDE that can be integrated into a state-space model as the state transition equation. 
On the other hand, solving such a dynamical model is non-trivial as it involves an infinite-dimensional SDE as the state transition equation; e.g., the filtering problem requires an infinite-dimensional Kalman filter or other approaches. Hence,  we resort to a finite-dimensional approximation of the infinite-dimensional SDE. Based on the first-order Euler approximation of the discretized SDE, the Gelerkin's method can be used to obtain a finite-dimensional approximation of the original infinite-dimensional SDE \citep{amsallem2014stability, benner2015survey}, which leads to a linear state-space model with a finite number of states. Then, leveraging the Karhunen–Loève (KL) expansion of the Gaussian noise, the space-time covariance of the approximated state-space model is derived, and the approximation error between the space-time covariance of the approximated and the original dynamical models can also be obtained.  

It is worth noting that, if the proposed approach above is adopted,  half of the state variables in the proposed dynamical model are directly related to the first-order time derivative of the original spatio-temporal process (the other half determines the original spatio-temporal process). This special structure makes the model an ideal choice for monitoring the first-order time derivative of the process over the spatial domain, directly through the posterior filtering distribution of these state variables. 
Note that, in \cite{liu2022statistical}, the authors also proposed a linear finite-state state-space model for stochastic advection-diffusion process with a white-in-time Gaussian noise (i.e., a first-order process). A key difference is that their state space model does not involve any state variables directly related to the first-order time derivative of the original spatio-temporal process, and is shown (in our numerical examples) to be much less effective in monitoring the first-order time derivative (of the original spatio-temporal process). We also note that, \cite{quick2015bayesian} developed the distribution theory for spatiotemporal gradients leveraging the limiting properties of GP in the space-time domain. In this paper, on the other hand, the estimate of the process and its temporal gradients are all performed in the frequency domain so that it is possible to leverage the hierarchical structure of a dynamical model and enable efficient detection of subtle changes of the temporal gradients at different frequency levels. 

\vspace{4pt}
Based on the discussions above, the main contribution of the paper is that it proposes a new dynamical model motivated from processes governed by second-order SPDEs, and demonstrates the performance of the proposed approach through comprehensive numerical investigations. The main results of the paper can be summarized as: 
\begin{itemize}
    \item A new dynamical model for spatio-temporal processes with an infinite-dimensional linear state-space representation where the state transition equation is an SDE. 
    \item The finite-dimensional approximation to the infinite-dimensional dynamical model that facilitates computation and parameter estimation. The first half of the state variables determine the spatio-temporal process, while the second half of the state variables are directly related to the first-order time derivative of the process---a structure that makes the proposed model especially effective in monitoring sudden changes of the process in the frequency domain. 
    \item The space-time covariance of the approximated dynamical model, and the quantification of the approximation error between the approximated and original processes. All results above are given for both complex- and real-valued processes often encountered in scientific or engineering applications. 
    \item Comprehensive numerical investigations that yield useful insights on the advantages of the proposed dynamical model. 
\end{itemize}

The remainder of this paper is organized as follows. 
Section \ref{sec:model} presents the proposed model. In particular, Section \ref{sec:SDE} describes the proposed infinite-dimensional SDE used as the state transition equation, motivating examples, and some fundamental model properties. Section \ref{sec:Dynamcal} describes how the statistical dynamical model can be obtained from the finite-dimensional approximation using the Galerkin's method, the space-time covariance, as well as the model for both the general complex-valued and real-valued processes. 
Finally, four numerical investigations are provided in Section \ref{sec:numerical}, and Section \ref{sec:conclusions} concludes the paper.

\section{The Proposed Model and Its Properties} \label{sec:model}

We first present the generic structure of the dynamical model considered in this paper. Let $x(\bm{s},t)$ be a continuous space-time process on the spatial domain $\mathbb{R}^d$ and temporal domain $\mathbb{T}$, and define $\bm{x}(\bm{s},t) = (x(\bm{s},t),\; \dot{x}(\bm{s},t))^T$, 
where $\dot{x}(\bm{s},t)$ denotes the time derivative of $x(\bm{s},t)$. Let $Y(\bm{s},t)$ be the observation process of $x(\bm{s},t)$, we consider an infinite-dimensional dynamical model in the form: 
\begin{subequations} \label{eq:state-space}
\begin{equation}
    d\bm{x}(\bm{s},t)  = \mathcal{\bm{A}}\bm{x}(\bm{s},t)dt + \bm{\bm{\bm{\bm{L}}}}dW(\bm{s},t) \label{eq:state-space1}
\end{equation}
\begin{equation}
    Y(\bm{s},t)  = \bm{H}\bm{x}(\bm{s},t) + \varepsilon^{\text{obs}}(\bm{s},t) \label{eq:state-space2}
\end{equation}
\end{subequations}
 where $\bm{L} = (0,\,1)^T$, $\bm{H}=(1,0)$, $W(\bm{s},t)$ is a white-in-time spatial Wiener process,  
$\bm{\mathcal{A}}$ is a $2 \times 2$ matrix of operators, and $\varepsilon^{\text{obs}}(t,\bm{s})$ is a spatially-colored and white-in-time Gaussian noise to capture observation errors. 
Note that, at any fixed $t$, the field $x(\bm{s},t)$ is an element of an infinite-dimensional Hilbert space, $x(\bm{s},t) \in \mathcal{H}(\mathbb{R}^d)$. Hence, the state equation (\ref{eq:state-space1}) is an \textit{infinite-dimensional} stochastic differential equation (SDE), which makes it a non-trivial task to solve the dynamical model (\ref{eq:state-space}). 

In the following sections, Section \ref{sec:SDE} proposes a generic infinite-dimensional SDE that governs the spatio-temporal process $\bm{x}(\bm{s},t)$. After that, we show how a statistical dynamical model can be constructed for processes governed by such a SDE (Section \ref{sec:Dynamcal}). In particular, we provide details on how the model is motivated, investigate some useful properties of the proposed model and discuss the connections between the proposed and existing models. 

\subsection{The Proposed Stochastic Process Governed by an Infinite-Dimensional SDE} \label{sec:SDE}


The dynamics of $\bm{x}(\bm{s},t)$ is governed by an infinite-dimensional SDE in the form of (\ref{eq:state-space1}), i.e., $d\bm{x}(\bm{s},t) = \bm{\mathcal{A}}\bm{x}(\bm{s},t)dt + \bm{L}dW(\bm{s},t)$. 
In this paper, we propose to consider the $2 \times 2$ operator matrix $\bm{\mathcal{A}}$ with the following proposed structure:
\begin{equation} \label{eq:A_matrix}
    \bm{\mathcal{A}} = \begin{bmatrix} 0 & 1 \\ -\mathcal{A}_0 & -\gamma \end{bmatrix}, \quad\quad \gamma \geq 0
\end{equation}
and $\mathcal{A}_0$ is an operator acting on $x(\bm{s},t) \in \mathcal{H}(\mathbb{R}^d)$ motivated by the advection-diffusion-decay process
\begin{equation} \label{eq:A0_def}
    \mathcal{A}_0 = \eta - \frac{1}{2}\nabla^2_{\bm{\Sigma}}\cdot + \nabla_{\bm{\mu}}\cdot 
\end{equation}
where the decay $\eta > 0$, and the two spatial diffusion and advection differential operators, acting on $x(\bm{s},t) \in \mathcal{H}(\mathbb{R}^d)$, are respectively chosen to be
\begin{equation} \label{eq:operators}
    \nabla^2_{\bm{\Sigma}}\cdot \;=\; \sum_{i=1}^{d} \sigma_i^2 \frac{\partial^2}{\partial s_i^2}\quad, \qquad \nabla_{\bm{\mu}}\cdot \;=\; \sum_{i=1}^{d} \mu_i \frac{\partial}{\partial s_i}\quad
\end{equation}
with $\bm{\Sigma} = \mathrm{diag}(\sigma_1^2,\ldots,\sigma_d^2)$, and  
$\bm{\mu} = (\mu_1,\ldots,\mu_d)^T$. 

It is important to note that the process governed by the proposed  (\ref{eq:state-space1}) can be re-written as a second-order stochastic process given by a stochastic partial differential equation (SPDE):
\begin{equation} \label{eq:2nd_order_SPDE}
    \ddot{x}(\bm{s},t) + \gamma\,\dot{x}(\bm{s},t) + \mathcal{A}_0\, x(\bm{s},t) = \varepsilon(\bm{s}, t)
\end{equation}
where $\varepsilon(\bm{s},t) \equiv \dot{W}(\bm{s},t)$ is the \textit{generalized} derivative of a Wiener process, i.e., a white-in-time spatially-colored Gaussian noise.  

\subsubsection{Some Motivating Examples}
The SPDE (\ref{eq:2nd_order_SPDE}) is a generic form that involves many physics process as its special cases. We provide some examples to demonstrate its flexibility: 
\begin{itemize}
    \item \textbf{The Mass-Spring-Damper system.} Consider a block with a unit mass $m=1$ attached to a spring with stiffness $k$ and a damper (a shock absorber) with coefficient $c$, and pulled by an external force $F(t)$. The behavior of this system is governed by Newton's Second Law, resulting in a second-order differential equation, $\ddot{\xi}+ c\dot{\xi} + k\xi= F(t)$, where $\xi$ represents displacement of the mass. Let $\bm{x} = (x,\dot{x})^T$, we obtain 
    \begin{equation} \label{eq:state-space1_spring}
    d\bm{x} = \begin{bmatrix} 0 & 1 \\ -k & -c \end{bmatrix}
    \bm{x}dt + \begin{bmatrix} 0 \\ 1 \end{bmatrix}F(t),
\end{equation}
   which falls into the generic form (\ref{eq:state-space1}) considered in this paper with a special $\mathcal{A}_0 = k$ and $\gamma= c$. 

    \item \textbf{The Stochastic Wave Equation.} 
    The most common form of the (damped) stochastic wave equation for a displacement $x(\bm{s},t)$ is given by 
    $\frac{\partial^2 x}{\partial t^2} - c^2 \nabla^2 x = \varepsilon(\bm{s},t)$, where $\nabla^2$ is the Laplacian operator and $\varepsilon(\bm{s},t)$ represents a Gaussian white noise. Let $\bm{x} = (x,\dot{x})^T$, we obtain 
    \begin{equation} \label{eq:SWE}
    d\bm{x} = \begin{bmatrix} 0 & 1 \\ -\frac{c^2}{2}\nabla^2 & -\gamma \end{bmatrix}
    \bm{x}dt + \begin{bmatrix} 0 \\ 1 \end{bmatrix}dW(\bm{s},t),
\end{equation}
which also falls into the generic form (\ref{eq:state-space1}) with $\mathcal{A}_0 = \frac{c^2}{2}\nabla^2$ being the Laplacian operator. 

    \item \textbf{The Power Grid Wave Equation.} Let $x(\bm{s},t)$ be the voltage, and the second-order power wave equation is given by 
    $LC \frac{\partial^2 x}{\partial t^2} + (RC+LG) \frac{\partial x}{\partial t} - \nabla^2 x = - RGV$, where $L$, $C$, $R$ and $G$ are respectively the inductance per unit length, shunt capacitance per unit length, resistance and shunt conductance. Let $\bm{x} = (x,\dot{x})^T$, we obtain 
    \begin{equation} \label{eq:Grid}
    d\bm{x} = \begin{bmatrix} 0 & 1 \\ \frac{1}{LC}\nabla^2 & -\frac{RC+LG}{LC} \end{bmatrix}
    \bm{x}dt + \begin{bmatrix} 0 \\ 1 \end{bmatrix}\frac{-RGV}{LC},
\end{equation}
    which is consistent with the generic form (\ref{eq:state-space1}) with $\mathcal{A}_0 = -\frac{1}{LC}\nabla^2$ and $\gamma = \frac{RC+LG}{LC}$. 
\end{itemize}

Many other examples can be found to demonstrate the flexibility of the model (\ref{eq:state-space1}) considered in this paper. Since the statistical dynamical model is constructed based on (\ref{eq:state-space1}) in this paper, it is expected that the proposed statistical model is also flexible to accommodate a range of applications. 

\subsubsection{Some Key Properties}
The solution of the proposed SDE (\ref{eq:state-space1}), or (\ref{eq:2nd_order_SPDE}) equivalently, generates a stationary spatio-temporal Gaussian Process (GP). 
Taking the Fourier transform of~(\ref{eq:2nd_order_SPDE}), on both sides, in both space and time yields the power spectral density of $x(\bm{s},t)$. Note that, 
the Fourier transform (in space) of the two spatial diffusion and advection differential operators in (\ref{eq:A0_def}) are respectively 
\begin{equation} \label{eq:verify_operators_1}
\mathcal{F}_s(\triangledown^2_{\bm{\Sigma}}\cdot)=\sum_{i=1}^{d}\sigma_i^2\mathcal{F}_s(\frac{\partial^2}{\partial s_i^2})=\sum_{i=1}^{d}u_i^2\sigma_i^2=-\bm{u}^T\bm{\Sigma} \bm{u} 
 \end{equation}
\begin{equation} \label{eq:verify_operators_2}
\mathcal{F}_s(\triangledown_{\bm{\mu}}\cdot)=\sum_{i}^{d}\mu_i\mathcal{F}_s(\frac{\partial}{\partial s_i})=i \sum_{i}^{d}  u_i \mu_i = i \bm{u}^T\bm{\mu} 
 \end{equation}
 
Then, the Fourier transform of (\ref{eq:2nd_order_SPDE}) yields
\begin{equation} \label{eq:freq_domain}
    (-v^2 + i\gamma v + \tilde{\mathcal{A}}_0(\bm{u}))\tilde{x}(\bm{u},v) = \tilde{\varepsilon}(\bm{u},v)
\end{equation}
where $\tilde{\mathcal{A}}_0(\bm{u}) = \eta + \tfrac{1}{2}\bm{u}^T \bm{\Sigma}\, \bm{u} + i\,\bm{u}^T \bm{\mu}$, and $\tilde{x}(\bm{u},v)$ and $\tilde{\varepsilon}(\bm{u},v)$ are the Fourier transform of $x(\bm{s}, t)$ and $\varepsilon(\bm{s}, t)$ respectively. Hence, in the frequency domain, it follows from (\ref{eq:freq_domain}) that $\tilde{x}(\bm{u},v)=\tilde{\varepsilon}(\bm{u},v)/(-v^2 + i\gamma v + \tilde{\mathcal{A}}_0(\bm{u}))$. 
In this paper, we let the spatial covariance of $\varepsilon(\bm{s}, t)$ be $Q(\bm{h})=(2\pi)^{-d/2}|\bm{\bm{\Phi}}|^{-1/2}\exp\left\{ -\bm{h}^T\bm{\bm{\Phi}}^{-1} \bm{h}/2 \right\}$ where $\bm{h}=\bm{s}-\bm{s}' \in \mathbb{R}^d$. Then, the power spectral density of $x(\bm{s}, t)$ is given by $S(\bm{u},v) = |\tilde{x}(\bm{u},v)|^2$, which is
\begin{equation} \label{eq:power_spectrum_simplified}
    S(\bm{u},v) =\frac{\exp(-\bm{u}^T\bm{\Phi}\,\bm{u}/2)}{(-v^2 + \eta + \tfrac{1}{2}\bm{u}^T\bm{\Sigma}\,\bm{u})^2 + (\gamma v + \bm{u}^T\bm{\mu})^2}
\end{equation}
where $\exp(-\bm{u}^T\bm{\Phi}\,\bm{u}/2)$ comes from the Fourier transform  $\mathcal{F}(Q(\bm{h}))$.





\noindent \textbf{Remark 1.}
    Using contour integration, the spatial power spectrum is given by
\begin{equation} \label{eq:spatial_spectrum}
\begin{aligned}
    S(\bm{u}) & = \int_{-\infty}^{\infty} S(\bm{u},v)\,dv \\ & = \frac{\pi\,\exp\!\bigl(-\bm{u}^T\bm{\Phi}\,\bm{u}/2\bigr)}{\gamma(\eta + \tfrac{1}{2}\bm{u}^T\bm{\Sigma}\,\bm{u} - \gamma^{-2}(\bm{u}^T\bm{\mu})^2)}.
\end{aligned}
\end{equation}
For this spatial spectrum to be well-defined and strictly positive, the denominator must be strictly positive for all spatial frequencies $\bm{u}$
\begin{equation} \label{eq:stationarity_condition}
    \eta + \tfrac{1}{2}\bm{u}^T\bm{\Sigma}\,\bm{u} > \frac{(\bm{u}^T\bm{\mu})^2}{\gamma^2} \quad \forall \bm{u}.
\end{equation}

This condition requires  $\bm{\Sigma}$ to be sufficiently large relative to $\bm{\mu}$, ensuring that the spatial propagation does not destabilize the process. If we view the operator (\ref{eq:A0_def}) as an advection-diffusion operator, condition~(\ref{eq:stationarity_condition}) implies  that the diffusive dissipation dominates over advective energy transport.

\vspace{4pt}
\noindent \textbf{Remark 2.}
    The behavior of the temporal covariance is determined by the poles of the spectrum, which correspond to the roots of the system's characteristic equation:
\begin{equation} \label{eq:characteristic_eq}
    \lambda^2 + \gamma \lambda + ( \eta + \tfrac{1}{2}\bm{u}^T\bm{\Sigma}\,\bm{u} + i\bm{u}^T\bm{\mu} ) = 0.
\end{equation}
The roots are given as $ -\frac{\gamma}{2} \pm \frac{1}{2}(\gamma^2 - 4(\eta + \tfrac{1}{2}\bm{u}^T\bm{\Sigma}\,\bm{u} + i\bm{u}^T\bm{\mu}))^{1/2}$. Notably, because the term inside the square root is complex and can have a negative real part at high spatial frequencies, the roots naturally consist of a real decay component and an imaginary oscillatory component. 
By applying the residue theorem, the temporal covariance in the mixed domain takes the form of an exponentially damped oscillation:
\begin{equation} \label{eq:damped_oscillation}
    C(\bm{u}, \Delta t) \;\propto\; e^{-\frac{\gamma}{2}|\Delta t|} \cos(\omega(\bm{u})|\Delta t| - \phi(\bm{u})),
\end{equation}
where the decay rate is strictly governed by $\gamma/2$, and $\omega(\bm{u})$ and $\phi(\bm{u})$ are the frequency and phase shift arising from the complex roots. The oscillatory temporal structure in~(\ref{eq:damped_oscillation}) arises when the spatial frequencies are sufficiently high, making this model uniquely suited for representing wave-like propagation phenomena in spatio-temporal data.


\vspace{4pt}
\noindent  \textbf{Remark 3 (differences from convolution-generated models).}
It is interesting to compare the power spectrum $S(\bm{u},v)$ in (\ref{eq:power_spectrum_simplified}) with that of the first-order convolution-generated process $Y(\bm{s},t)$ proposed in \cite{Brown2000}, which is also the solution from a linear stochastic advection-diffusion equation with a Gaussian noise \citep{Sigrist2015, liu2022statistical}. The latter has power spectrum
\begin{equation} \label{eq:convolution_spectrum}
    S_C(\bm{u},v) = \frac{\exp\!(-\bm{u}^T\bm{\Phi}\,\bm{u}/2)}{(\eta + \tfrac{1}{2}\bm{u}^T\bm{\Sigma}\,\bm{u})^2 + (v + \bm{u}^T\bm{\mu})^2}.
\end{equation}

Comparing~(\ref{eq:power_spectrum_simplified}) with~(\ref{eq:convolution_spectrum}), the key difference is the presence of the \emph{inertial term}, $-v^2$, and the damping parameter $\gamma$ in the second-order model. In the spectrum~(\ref{eq:convolution_spectrum}), the denominator is quadratic in $v$ with a positive leading coefficient, ensuring a single spectral peak at $v = -\bm{u}^T\bm{\mu}$. The corresponding temporal covariance exhibits an envelope that is a strictly monotonic exponential decay $e^{-\alpha|\Delta t|}$ (with a phase shift representing pure spatial translation), and well represents smooth, diffusion-like dynamics. In contrast, the second-order spectrum~(\ref{eq:power_spectrum_simplified}) has a denominator that is quartic in $v$, which can produce two spectral peaks for each spatial frequency. This richer spectral structure is what gives rise to the damped oscillatory temporal covariance in~(\ref{eq:damped_oscillation}).

\vspace{4pt}
\noindent \textbf{Remark 4. asymptotic connection to first-order model}
It is worth noting that the second-order model strictly generalizes the classical first-order model. In physics, this connection is formally known as the \emph{overdamped limit} . If the damping parameter $\gamma$ becomes sufficiently large, the frictional force $\gamma\dot{x}$ dominates the inertial term $\ddot{x}$. Under this limit, the system's momentum is entirely suppressed, and the second-order SPDE asymptotically collapses back to the first-order advection-diffusion equation. 
Interestingly, if one were to extend the scalar damping $\mathcal{A}_1 = \gamma$ to a differential operator (e.g., stiffness-proportional damping $\mathcal{A}_1 = \gamma_0 - \gamma_1 \nabla^2$), the damping ratio would scale quadratically with the spatial frequency $\bm{u}$. Under such a viscoelastic regime, the second-order system can perfectly and universally recover the exact first-order diffusion behavior across all spatial scales. However, to explicitly showcase the wave-propagating and oscillatory nature distinct from purely diffusive models, we maintain a constant scalar $\gamma$ in our primary framework.



\subsection{The Statistical Dynamical Model} \label{sec:Dynamcal}

\subsubsection{A Finite-Dimensional Approximation using the Galerkin's Method.} \label{sec:approximation}

Solving the dynamical model (\ref{eq:state-space1}) is non-trivial because it involves an infinite-dimensional SDE (\ref{eq:state-space1}) as the state equation; for example, the filtering problem may require an infinite-dimensional Kalman Filter. Hence, we resort to a finite-dimensional approximation obtained as follows. 

We start with the discretized version of the SDE (\ref{eq:state-space1}) as follows,
\begin{equation}\label{eq:discrete}
    \bm{x}(\bm{s},t_{\bm{k}}) = A(\delta)\bm{x}(\bm{s},t_{k-1}) + \bm{e}(\bm{s}), \quad k=1,2,\cdots, K
\end{equation}
where $\delta=t_{\bm{k}}-t_{k-1}$, $A(\delta)=e^{\mathcal{A}\delta}$, $\bm{e}(\bm{s})=\int_{0}^{\delta}A(\delta-\tau)\bm{L}W(\bm{s},\tau)d\tau$ is a spatial GP denoted by $GP(0,\bm{\Sigma}_e)$, and
\begin{equation} \label{eq:Sigma_e_1}
    \bm{\Sigma}_e = Q\int_{0}^{\delta}A(\delta-\tau)\bm{L}\bm{L}^T A^*(\delta-\tau)d\tau. 
\end{equation}
where $Q$ is defined above (\ref{eq:power_spectrum_simplified}).
Because $A$ is the exponential of an operator, we adopt, for a small $\delta$, the first-order Euler approximation of (\ref{eq:discrete}) and obtain
\begin{equation} \label{eq:Euler}
    \bm{x}(\bm{s},t_{\bm{k}}) = \bm{x}(\bm{s},t_{k-1}) + \mathcal{A}\delta \bm{x}(\bm{s},t_{k-1}) + \bm{e}(\bm{s},t_{k-1}). 
\end{equation}

Note that, although (\ref{eq:Euler}) is still an infinite-dimensional problem because $x(\bm{s},t),\dot{x}(\bm{s},t)\in\mathcal{H}(\mathbb{R}^d)$, the Galerkin's method can be used to obtain a finite-dimensional approximation of (\ref{eq:Euler}). 
Let $\phi_j(\bm{s})=e^{i\bm{\omega}_j^Ts}$ be the Fourier eigenfunction of $\mathcal{A}_0$ with the spatial frequencies $\bm{\omega}_j=(\bm{\omega}_{j,1},\bm{\omega}_{j,2}, \cdots, \bm{\omega}_{j,d})^T$. Performing the eigenfunction decomposition of the operator $\mathcal{A}_0$, we obtain
\begin{equation}\label{equation:eigen_function}
\begin{aligned}
 & \mathcal{A}_0\phi_j(\bm{s})  = \eta\,\phi_j(\bm{s}) - \frac{1}{2}\triangledown^2_{\bm{\Sigma}}\cdot \phi_j(\bm{s}) + \triangledown_{\bm{\mu}}\cdot \phi_j(\bm{s}) \\
& \quad = \Big(\eta - \frac{1}{2} \sum_{i=1}^d \sigma_i^2\,\omega_{j,i}^2
        + i \sum_{i=1}^d \mu_i\,\omega_{j,i}\Big)\,\phi_j(\bm{s})
\; \\ & \quad \equiv\; \lambda_j\,\phi_j(\bm{s})
\end{aligned}
\end{equation}
where $\lambda_j$ is the eigenvalue of the operator $\mathcal{A}_0$. 
Then, let $x(\bm{s},t)=\sum_{j=1}^{J}\phi_j(\bm{s})\alpha_j(t)\equiv \bm{\phi}^T(\bm{s})\bm{\alpha}(t)$ and $\dot{x}(\bm{s},t)=\sum_{j=1}^{J}\phi_j(\bm{s})\beta_j(t)\equiv \bm{\phi}^T(\bm{s})\bm{\beta}(t)$, (\ref{eq:Euler}) is re-written as
\begin{equation} \label{eq:discrete_17}
    \begin{aligned} 
& \begin{bmatrix}
\bm{\phi}^T(\bm{s})\alpha(t_{\bm{k}}) \\ \bm{\phi}^T(\bm{s})\beta(t_{\bm{k}})
\end{bmatrix}   = \begin{bmatrix}
\bm{\phi}^T(\bm{s})\alpha(t_{k-1}) \\ \bm{\phi}^T(\bm{s})\beta(t_{k-1})
\end{bmatrix} \\ & \quad\quad + 
\begin{bmatrix}
\bm{\phi}^T(\bm{s})\beta(t_{k-1}) \\  -\mathcal{A}_0\bm{\phi}^T(\bm{s})\alpha(t_{k-1}) - \gamma\bm{\phi}^T(\bm{s})\beta(t_{k-1})
\end{bmatrix} \delta \\
& \quad\quad + \bm{e}(\bm{s},t_{k-1}).
\end{aligned}
\end{equation}

Following the eigenfunction decomposition $\mathcal{A}_0\bm{\phi}_\jmath=\lambda_\jmath \bm{\phi}_\jmath$, and recall the fact that the Fourier eigenfunctions are orthogonal (i.e., $\int \phi_j \phi^*_\jmath ds$ is a constant when $j = \jmath$, and $\int \phi_j \phi^*_\jmath ds=0$ otherwise), multiplying both side by $\phi_\jmath (\bm{s})$ and then integrate both sides over the spatial domain yields the following for any $\bm{\omega}_\jmath$ ($\jmath=1,2,\cdots,J$)
\begin{equation} \label{eq:discrete_18}
\begin{bmatrix}
\alpha_\jmath(t_{\bm{k}}) \\ \beta_\jmath(t_{\bm{k}})
\end{bmatrix} = \begin{bmatrix}
1 & \delta\\ -\lambda_\jmath \delta & 1-\gamma\delta
\end{bmatrix}  \begin{bmatrix}
\alpha_\jmath(t_{k-1}) \\ \beta_\jmath(t_{k-1})
\end{bmatrix} 
+ \bm{v}_\jmath(t_{k-1})
\end{equation}
where $\bm{v}_\jmath$ is a random vector and we leave its derivation to Section \ref{sec:covariance}. Compared with (\ref{eq:Euler}), (\ref{eq:discrete_18}) describes the (approximated) finite-state temporal evolution at any frequency $\bm{\omega}_\jmath$.  

\subsubsection{The Space-Time Covariance.} \label{sec:covariance}


In this section, we obtain the expression of $\bm{v}_\jmath(t)$ in (\ref{eq:discrete_18}). 
Firstly, note that $\bm{e}(\bm{s})=\int_{0}^{\delta}A(\delta-\tau)\bm{L}W(\bm{s},\tau)d\tau$ is a spatial GP, i.e., $GP(0,\bm{\Sigma}_e)$ with $\bm{\Sigma}_e = Q\int_{0}^{\delta}A(\delta-\tau)\bm{L}\bm{L}^T A^*(\delta-\tau)d\tau$ in the discretized SDE  (\ref{eq:discrete}). When the first-order Euler
approximation is adopted, we have
\begin{equation}
 \begin{aligned}
    A(\delta-\tau) &  =e^{\mathcal{A}(\delta-\tau)} \approx \bm{I} + \mathcal{A}(\delta-\tau) \\ & = \bm{I} +(\delta-\tau) \begin{bmatrix}
0 & 1\\ -\mathcal{A}_0 & -\gamma \end{bmatrix}  
\\ & =  \begin{bmatrix}
1 & \delta-\tau\\ -\mathcal{A}_0(\delta-\tau) & 1-\gamma \delta+\gamma \tau \end{bmatrix} 
\end{aligned}   
\end{equation}
and $A(\delta-\tau)\bm{\bm{\bm{\bm{L}}}}  \approx (\delta-\tau, 1-\gamma \delta+\gamma \tau)^T$, which is substituted into (\ref{eq:Sigma_e_1}) to yield
\begin{equation} \label{eq:Sigma_e}
  \bm{\Sigma}_e  \approx 
   Q \begin{bmatrix}
m_1 & m_{12}\\ m_{12} & m_2 
\end{bmatrix}   
\end{equation}
where $m_{1} = \int_{0}^{\delta} (\delta-\tau)^2\,d\tau = \frac{1}{3}\delta^3$, $m_{12} = \int_{0}^{\delta} (\delta-\tau)(1-\gamma\delta+\gamma\tau)\,d\tau = \frac{1}{2}\delta^2 - \frac{1}{3}\gamma \delta^3$, and $m_{2} = \int_{0}^{\delta} (1-\gamma\delta+\gamma\tau)^2\,d\tau = \delta - \gamma\delta^2 + \frac{5}{6}\gamma^2\delta^3$. 

The equation above provides a much simpler form of the covariance of $\bm{e}(\bm{s})$ based on the first-order Euler approximation. 
However, to obtain the finite-dimensional approximation of (\ref{eq:discrete_18}) using the Galerkin's method above, we need a spectrum representation of $\bm{e}(\bm{s},t)$. The following Proposition 1 allows us to well approximate $\bm{e}(\bm{s},t)$ by $\tilde{\bm{e}}(\bm{s},t)$, and a spectrum representation is available for the latter.  
\begin{proposition}
  Let $\tilde{\bm{e}}(\bm{s},t)=(\sqrt{m_1\delta^{-1}},\sqrt{m_2\delta^{-1}})^TdW(\bm{s},t)$, and $W(\bm{s},t)$ is a Wiener process with $\mathbb{E}[dW(\bm{s},\tau)\,dW(\bm{s}',\tau')]=Q(\bm{h})\delta_{\tau,\tau'}$ where $\delta_{\tau,\tau'}$ is the Kronecker delta. Then, we have the following results:
  
  1). $\tilde{\bm{e}}(\bm{s},t)$ admits a spectrum representation
\begin{equation}\label{eq:spectrum_e}
      \tilde{\bm{e}}(\bm{s},t) \approx  (\sqrt{m_1},\sqrt{m_2})^T \sum_{j=1}^{J}\sqrt{\kappa_j}\phi_j(\bm{s})Z_j
  \end{equation}
  where $\bm{\phi}_j$ is the Fourier eigenfunction, $\kappa_j=\exp\{-\bm{\omega}_j^T\bm{\bm{\Phi}}\bm{\omega}_j/2\}$, and $Z_j$, $j=1,\cdots,J$, are independent standard Gaussian random variables. 
  
  2). $\tilde{\bm{e}}(\bm{s},t) \sim N(0,\bm{\Sigma}_{\tilde{\bm{e}}})$, and 
  \begin{equation} \label{eq:error}
\bm{\Sigma}_{\tilde{\bm{e}}}-\bm{\Sigma}_{\bm{e}} = \begin{bmatrix}
0 &  O(\delta^2)\\ O(\delta^2) & 0 \end{bmatrix} .
  \end{equation}
\end{proposition}

\vspace{8pt}
The spectrum representation of $\tilde{\bm{e}}(t)$ in (\ref{eq:spectrum_e}) directly comes from the Karhunen–Loève (KL) expansion of a GP, $dW(\bm{s},t)=\sqrt{\delta}\sum_{j=1}^{\infty}\sqrt{\kappa_j}\phi_j(\bm{s})Z_j\sim GP(0,Q(\bm{h})\delta)$. It immediately follows that $\tilde{\bm{e}}(t) \approx  (\sqrt{m_1},\sqrt{m_2})^T \sum_{j=1}^{J}\sqrt{\kappa_j}\phi_j(\bm{s})Z_j$. In addition, note that $\mathbb{E}\{dW(\bm{s},t)dW(\bm{s}',t)\}=Q(\bm{h})\delta$. From the KL expansion of $dW(\bm{s},t)$, we also have $\mathbb{E}\{dW(\bm{s},t)dW(\bm{s}',t)\}=\delta\sum_{j=1}^{\infty}\kappa_j \bm{\phi}(\bm{s})\bm{\phi}^T(\bm{s}')=\delta\sum_{j=1}^{\infty}\kappa_j e^{i\bm{\omega}_j^T \bm{h}}$. Hence, $Q(\bm{h})=\sum_{j=1}^{\infty}\kappa_j e^{i\bm{\omega}_j^T \bm{h}}$, and $\kappa_j$ is found from the inverse Fourier transform of a Gaussian density function $Q(\bm{h})$, given in (\ref{eq:Q}), as $\kappa_j=\exp\{-\bm{\omega}_j^T\bm{\bm{\Phi}}\bm{\omega}_j/2\}$. 

To obtain the approximation error $\bm{\Sigma}_{\tilde{\bm{e}}}-\bm{\Sigma}_{\bm{e}}$ shown in (\ref{eq:error}), it follows from $\tilde{\bm{e}}(\bm{s},t)=(\sqrt{m_1\delta^{-1}},\sqrt{m_2\delta^{-1}})^TdW(\bm{s},t)$ that 
\begin{equation}\label{eq:Sigma_e_tilde}
\begin{aligned}
  \bm{\Sigma}_{\tilde{e}} =  Q\begin{bmatrix}
m_1 & \sqrt{m_1m_2}\\ \sqrt{m_1m_2} & m_2 
\end{bmatrix}. 
\end{aligned}
\end{equation}

Comparing $\bm{\Sigma}_e$ in (\ref{eq:Sigma_e}) and $\bm{\Sigma}_{\tilde{e}}$ in (\ref{eq:Sigma_e_tilde}), it is seen that the elements on the diagonal line are exactly the same, while the difference of the off-diagonal entries is
\begin{equation}
\begin{aligned}
& \sqrt{m_1m_2} - m_{12} \\ & \quad = 
\delta^2\sqrt{\frac{1}{3} -\frac{\gamma\delta}{3} + \frac{5\gamma^2\delta^2}{18}} - \delta^2\left( \frac{1}{2} - \frac{\gamma\delta}{3} \right) 
= O(\delta^2).    
\end{aligned}
\end{equation}

\subsection{The Model for General Complex-Valued Processes}
\label{sec:general}

Based on the results above, this section presents the dynamical model as follows: 
\begin{equation} \label{eq:discrete_17b}
\begin{aligned}
  &  \begin{bmatrix}
\bm{\phi}^T(\bm{s})\alpha(t_{\bm{k}}) \\ \bm{\phi}^T(\bm{s})\beta(t_{\bm{k}})
\end{bmatrix}  = \begin{bmatrix}
\bm{\phi}^T(\bm{s})\alpha(t_{k-1}) \\ \bm{\phi}^T(\bm{s})\beta(t_{k-1})
\end{bmatrix} \\ & \quad \quad + 
\begin{bmatrix}
\bm{\phi}^T(\bm{s})\beta(t_{k-1})\\ -\mathcal{A}_0\bm{\phi}^T(\bm{s})\alpha(t_{k-1}) - \gamma\bm{\phi}^T(\bm{s})\beta(t_{k-1})
\end{bmatrix} \delta \\
& \quad \quad + \begin{bmatrix} \sqrt{m_1} \sum_{j=1}^{J}\sqrt{\kappa_j}\phi_j(\bm{s})Z_j \\ \sqrt{m_2} \sum_{j=1}^{J}\sqrt{\kappa_j}\phi_j(\bm{s})Z_j\end{bmatrix}, 
\end{aligned}
\end{equation}
and equation (\ref{eq:discrete_18}) is also revised accordingly using the Galerkin projection
\begin{equation} \label{eq:discrete_18b}
\begin{aligned} 
\begin{bmatrix}
\alpha_\jmath(t_{\bm{k}}) \\ \beta_\jmath(t_{\bm{k}})
\end{bmatrix} & = \begin{bmatrix}
1 & \delta\\ -\lambda_\jmath \delta & 1-\gamma \delta
\end{bmatrix}  \begin{bmatrix}
\alpha_\jmath(t_{k-1}) \\ \beta_\jmath(t_{k-1})
\end{bmatrix} \\ & 
+ \begin{bmatrix}
\sqrt{m_1\kappa_\jmath} Z_\jmath \\\sqrt{m_2\kappa_\jmath} Z_\jmath
\end{bmatrix}.
\end{aligned}
\end{equation}

Finally, let $\bm{\theta}_j(t)=(\alpha(t), \beta_j(t))^T$, $\bm{\theta}(t)=(\bm{\theta}_1^T(t),\cdots,\bm{\theta}_J^T(t))^T$, 
$\bm{G}_j=\bigl(\begin{smallmatrix}
1 & \delta \\
-\lambda_j\delta & 1-\gamma\delta \\
\end{smallmatrix}\bigr)$, $\bm{G}=\text{diag}(\bm{G}_1,\bm{G}_2, \cdots, \bm{G}_J)$, $\bm{v}_\jmath=(\sqrt{m_1\kappa_\jmath} Z_\jmath, \sqrt{m_2\kappa_\jmath} Z_\jmath)^T$, and $\bm{v}=(\bm{v}_1^T,\cdots,\bm{v}_J^T)^T$, we obtain the finite-dimensional approximation of the infinite-dimensional dynamical model (\ref{eq:state-space}):
\begin{subequations} \label{eq:solution}
\begin{equation}
  \bm{\theta}(t_{\bm{k}})=\bm{G}\bm{\theta}(t_{k-1})+\bm{v}(t_{\bm{k}}) 
\end{equation}
\begin{equation}
  \Tilde{\bm{Y}}(\bm{s},t_{\bm{k}}) = \bm{H}\bm{\bm{\phi}}^T(\bm{s})\bm{\theta}(t_{\bm{k}}) + \bm{\varepsilon}(\bm{s},t). 
\end{equation}
\end{subequations}

Note that, the state vector $\bm{\theta}(t)$ now has a finite dimension of $J$.

\subsubsection{Real-Valued Processes.} \label{sec:real_value}

The discussions above apply to general processes that take complex values. In  scientific and engineering applications, one typically deals with real-valued processes. This section presents the special model formulation tailored for real-valued processes. 

Consider the spectral representation of a real-valued process and its time derivative as follows \citep{liu2022statistical}
\begin{equation}\label{equation:real_valued_0}
\begin{aligned}
    x(\bm{s},t) = & \sum_{{\bm{k}} \in \Omega_1}\alpha_{\bm{k}}^R(t)f_{\bm{k}}^R(\bm{s}) \\& + 2\sum_{{\bm{k}} \in \Omega_2}(\alpha_{\bm{k}}^R(t)f_{\bm{k}}^R(\bm{s})+\alpha_{\bm{k}}^I(t)f_{\bm{k}}^I(\bm{s}))
\end{aligned}
\end{equation}
\begin{equation}
\begin{aligned}
    \dot{x}(\bm{s},t_{n}) = & \sum_{\bm{k} \in \Omega_1}\beta_{\bm{k}}^R(t_n)f_{\bm{k}}^R(\bm{s}) \\&+ 2\sum_{\bm{k} \in \Omega_2}(\beta_{\bm{k}}^R(t_n)f_{\bm{k}}^R(\bm{s})+\beta_{\bm{k}}^I(t_n)f_{\bm{k}}^I(\bm{s}))
    \end{aligned}
\end{equation}
where $\alpha_{\bm{k}}^R(t)$,  $\alpha_{\bm{k}}^I(t)$, $\beta_{\bm{k}}^R(t)$ and $\beta_{\bm{k}}^I(t)$ are temporally-varying coefficients, $f_{\bm{k}}^R = \cos(2\pi \bm{k}^T\bm{s})$, $f_{\bm{k}}^I = \sin(2\pi \bm{k}^T\bm{s})$, $\Omega_1 = \{(0,0), (0,\frac{N_2}{2}), (\frac{N_1}{2},0), (\frac{N_1}{2},\frac{N_2}{2})\}$, and $\Omega_2 = \{(k_1,k_2); k_1=0,1,\ldots,\frac{N_1}{2},  k_2 = 0,1,\ldots,\frac{N_2}{2}\}  
         \cup \{(k_1,k_2); k_1= 1,\ldots,\frac{N_1}{2}-1,  k_2 = -1,\ldots,-\frac{N_2}{2}+1\} \setminus \Omega_1$.

Substituting the spectral representations above into (\ref{eq:Euler}) yields:
\begin{equation} 
\begin{bmatrix}
x(\bm{s},t_{k})\\
\dot{x}(\bm{s},t_{k})
\end{bmatrix} = 
\begin{bmatrix}
x(\bm{s},t_{k-1})\\
\dot{x}(\bm{s},t_{k-1})
\end{bmatrix}  + 
\begin{bmatrix}
\dot{x}(\bm{s},t_{k-1}) \\
-\mathcal{A}_0x(\bm{s},t_{k-1})- \gamma \dot{x}(\bm{s},t_{k-1})
\end{bmatrix} \delta
 + \bm{e}(\bm{s},t_{k-1})
\end{equation}
where $\mathcal{A}_0x(\bm{s},t)  = \sum_{\bm{k} \in \Omega_1}\alpha_{\bm{k}}^R(t) \cdot \mathcal{A}_0f_{\bm{k}}^R(\bm{s})+ 2\sum_{\bm{k} \in \Omega_2}( \alpha_{\bm{k}}^R(t) \cdot \mathcal{A}_0f_{\bm{k}}^R(\bm{s})+\alpha_{\bm{k}}^I(t) \cdot \mathcal{A}_0f_{\bm{k}}^I(\bm{s}))$
with $\mathcal{A}_0 f_{\bm{k}}^R(\bm{s})=a(\bm{k})f_{\bm{k}}^R(\bm{s}) - b(\bm{k})f_{\bm{k}}^I(\bm{s})$, $\mathcal{A}_0 f_{\bm{k}}^I(\bm{s})=b(\bm{k})f_{\bm{k}}^R(\bm{s}) + a(\bm{k})f_{\bm{k}}^I(\bm{s})$, $a(\bm{k})=\eta + 2\pi^2 \bm{k}^T\bm{\Sigma}\bm{k}$ and $b(\bm{k})=2\pi \bm{k}^T \bm{\mu}$. 

Following the same idea behind Proposition 1, $\bm{e}(\bm{s},t)=(\sqrt{m_1\delta^{-1}},\sqrt{m_2\delta^{-1}})^TdW(\bm{s},t)$. Here, the Wiener process $W(\bm{s},t)$ with $\mathbb{E}[dW(\bm{s},t)\,dW(\bm{s}',t')]=Q(\bm{h})\delta_{t,t'}$ can be represented by the real-valued version of the KL expansion:
\begin{equation}
\delta^{-1/2} dW(\bm{s},t)  =\sum_{{\bm{k}} \in \Omega_1}\sqrt{\zeta_{\bm{k}}^{1,R}}f_{\bm{k}}^R(\bm{s})Z^{1,R} + 2\sum_{{\bm{k}} \in \Omega_2}\sqrt{\zeta_{\bm{k}}^{2,R}}f_{\bm{k}}^R(\bm{s})Z^{2,R}+\sqrt{\zeta_{\bm{k}}^{2,I}}f_{\bm{k}}^I(\bm{s})Z^{2,I} 
\end{equation}
where $Z^{1,R}$, $Z^{2,R}$ and $Z^{2,I}$ are independent standard Gaussian random variables, and $\zeta_{\bm{k}}^{1,R}$, $\zeta_{\bm{k}}^{2,R}$ and $\zeta_{\bm{k}}^{2,I}$ are obtained from the spectral decomposition (Fourier transform) of $Q(\bm{h})$
\begin{equation}
    Q(\bm{h}) = \sum_{{\bm{k}} \in \Omega_1}\zeta_{\bm{k}}^{1,R}f_{\bm{k}}^R(\bm{h}) + 2\sum_{{\bm{k}} \in \Omega_2}\zeta_{\bm{k}}^{2,R}f_{\bm{k}}^R(\bm{h})+\zeta_{\bm{k}}^{2,I}(t)f_{\bm{k}}^I(\bm{h}). 
\end{equation}

Similar to how (\ref{eq:discrete_18b}) is obtained, the Galerkin's method yields: 

$\bullet$ For any $\bm{k} \in \Omega_1$
\begin{equation}\label{eq:real_transition_1}
   \begin{bmatrix}
    \alpha_{\bm{k}}^R(t_n) \\ \beta_{\bm{k}}^R(t_n)
    \end{bmatrix}  = \bm{G_1}  \begin{bmatrix}
    \alpha_{\bm{k}}^R(t_{n-1}) \\ \beta_{\bm{k}}^R(t_{n-1})
    \end{bmatrix} 
    + \begin{bmatrix}
\sqrt{m_1\zeta_{k}^{1,R}}Z^{1,R}\\ \sqrt{m_2\zeta_{k}^{1,R}}Z^{1,R}
    \end{bmatrix} 
\end{equation}

$\bullet$ For any $\bm{k} \in \Omega_2$
\begin{equation}
\label{eq:real_transition_2}
    \begin{bmatrix}
    \alpha_{\bm{k}}^R(t_n) \\ \beta_{\bm{k}}^R(t_n) \\ \alpha_{\bm{k}}^I(t_n) \\ \beta_{\bm{k}}^I(t_n)
    \end{bmatrix}  = \begin{bmatrix}
    \bm{G_1} & \bm{G_2}\\ -\bm{G_2} & \bm{G_1}
    \end{bmatrix}  \begin{bmatrix}
    \alpha_{\bm{k}}^R(t_{n-1}) \\ \beta_{\bm{k}}^R(t_{n-1}) \\ \alpha_{\bm{k}}^I(t_{n-1}) \\ \beta_{\bm{k}}^I(t_{n-1})
    \end{bmatrix} 
     + \begin{bmatrix}
\sqrt{m_1\zeta_{k}^{2,R}}Z^{2,R}\\ \sqrt{m_2\zeta_{k}^{2,R}}Z^{2,R}\\
\sqrt{m_1\zeta_{k}^{2,I}}Z^{2,I}\\ \sqrt{m_2\zeta_{k}^{2,I}}Z^{2,I}\\
    \end{bmatrix} 
\end{equation}
where
\begin{equation}
    \bm{G_1} = \begin{bmatrix}
    1 & \delta\\ -a(\bm{k})\delta& 1-\gamma\delta
    \end{bmatrix}, \quad
     \bm{G_2} = \begin{bmatrix}
    0 & 0\\ -b(\bm{k})\delta & 0
    \end{bmatrix} 
\end{equation}

Given the state transition equations (\ref{eq:real_transition_1}) and (\ref{eq:real_transition_2}), one can immediately obtain the finite-dimensional approximation, in the same form as (\ref{eq:solution}), to a real-valued dynamical model (\ref{eq:state-space}).

\section{Numerical Examples} \label{sec:numerical}

In this section, four numerical examples are provided to illustrate the application of the proposed model. These examples utilize data generated from 2D (seismic) wave equation, advection-diffusion equation, and a real-world atmospheric aerosol propagation process. 

\subsection{Experiment I: 2D Damped Wave Propagation}
\textbf{Data and Experiment Setup.} In Example I, we consider the modeling of data generated from a 2D damped wave equation
\begin{equation} \label{eq:wave_example1}
\ddot{x} + \gamma\dot{x} + \eta\, x = c^2\nabla^2 x + Q + \varepsilon.
\end{equation}
This PDE is solved on a $100\times 100$ grid, $[0,10]^2$, with uniform wave speed $c=1.4$, damping $\gamma=0.22$, and decay $\eta=0.015$. A number of 10 wave sources of varying amplitude (25--50) and frequency (0.5--3.2\,Hz) are placed over the spatial domain. A Gaussian error with a standard deviation of 0.05 is added to the observations. 

Data are simulated for 650 time steps with $\Delta t=0.012$s. All model parameters are estimated, using data in the first 350 steps (i.e., 4.2s), from the standard Maximum Likelihood Estimation (MLE) for statistical dynamical models \citep{west1997bayesian}. 
At time step $T_{\mathrm{change}}=450$ ($t=5.40$s), two anomalous sources of amplitude 350 and frequencies 1.8 and 1.4\,Hz are injected at locations $(7.2,1.5)$ and $(2.2,7.0)$.

\begin{figure}[h!]
    \centering
    \includegraphics[width=0.9\linewidth]{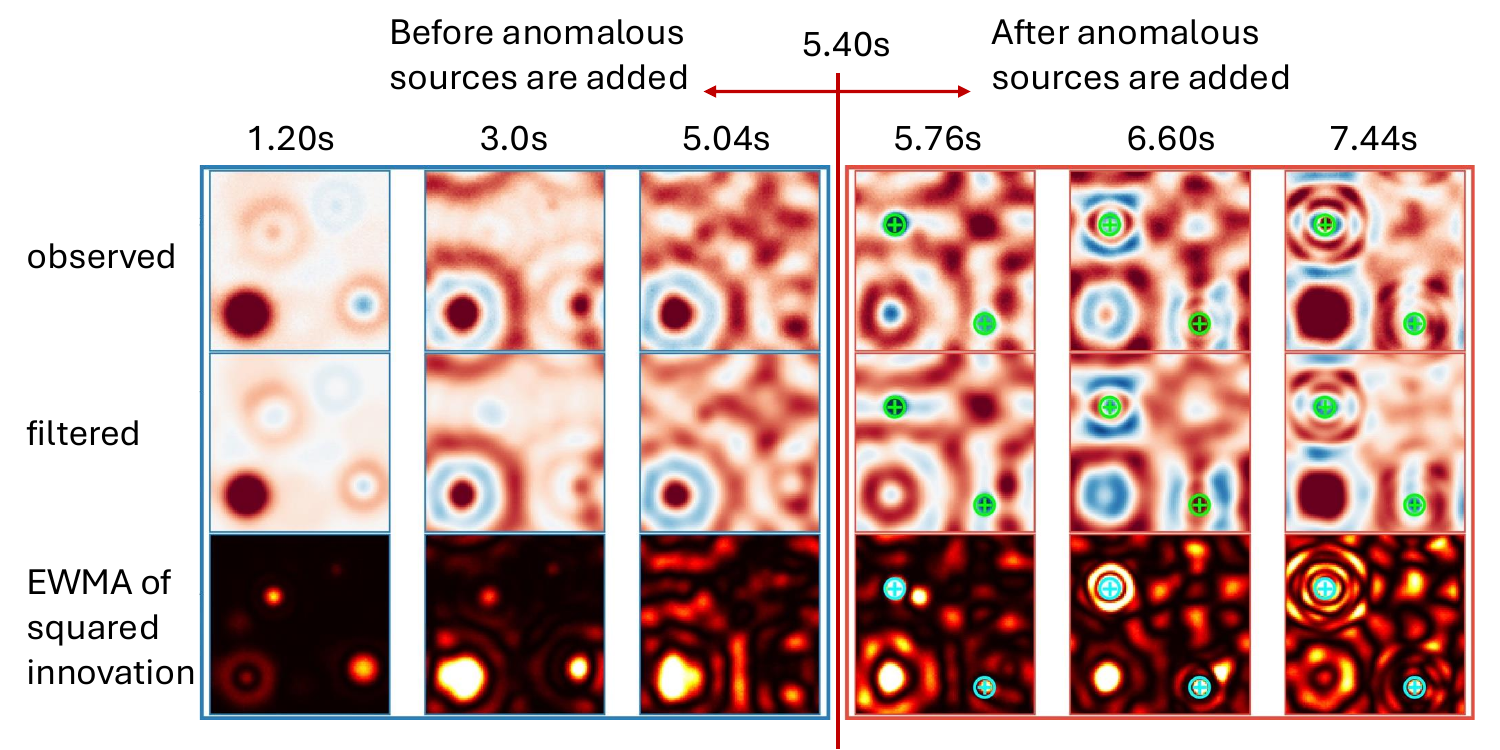}
    \caption{Observed (top) and filtered (middle) wave propagation, and the EWMA of the squared innovation (bottom) with the circles indicating the locations where anomalies are introduced.}
    \label{fig:kf_panels_exp1}
\end{figure}
\color{black}
\textbf{Results.} 
Figure \ref{fig:kf_panels_exp1} shows both the observed (top row) and filtered (middle row) processes at 6 chosen time steps: 1.2s, 3s, 5.04s, 5.76s, 6.6s and 7.44s. The filtered process is reconstructed through the inverse Fourier transform from the filtered $\bm{\theta}$ using (\ref{eq:solution}), and well captures the wave propagation process. Recall that the model parameters are estimated using data in the first 4.2s, and the two new sources (i.e., anomalies) are introduced at 5.4s. To demonstrate the anomaly detection capabilities of the proposed model, the one-step prediction error $v_t(i,j) = y_t(i,j) - \bm{H}\hat{\bm{x}}_{t|t-1}(i,j)$
is calculated for each pixel $(i,j)$. 
These residuals are white noise with a mean of zero before the two anomalies are introduced at $t=5.40$s.
The bottom row of Figure \ref{fig:kf_panels_exp1} shows the Exponentially Weighted Moving Average (EWMA) of the squared innovation,
$S_t(i,j) = \lambda\, v_t^2(i,j) + (1-\lambda)\, S_{t-1}(i,j)$, where $S_0 = 0$ and $\lambda=0.15$ in this numerical example.  
It is seen from this figure that the EWMA of the innovation energy remains spatially diffuse prior to the injection new sources. After $t=5.40$s, the two anomalous sources produce localized hotspots in the EWMA map that grow progressively stronger (bottom-right panels in Figure \ref{fig:kf_panels_exp1}), while the filter continues to track the overall wave field.

\vspace{-6pt}
\begin{figure}[h!]
    \centering
\includegraphics[width=1\linewidth]{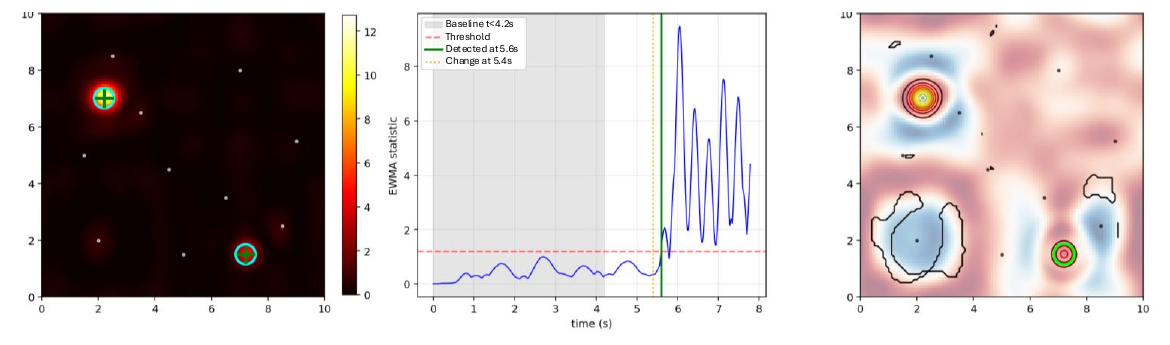}
\vspace{-24pt}
    \caption{EMWA of the squared innovation after and before the additional sources have been introduced (left), trajectory of $M_t$ (middle), contours on a post-injection wave field snapshot (right).}
    \label{fig:blind_exp1}
\end{figure}

\vspace{-12pt}
To better visualize the detected anomalies, a simple procedure is adopted as follows. At each time step, we obtain the maximum EWMA over the spatial domain, $M_t = \max_{i,j}\, S_t(i,j)$, and compare it to the maximum EWMA obtained during the model training phase before 4.2s. Then, an alarm is triggered when $M_t$ exceeds $k M_{\mathrm{base}}$, where $k$ is chosen to be 1.2 in this example for illustrative purposes. 
The left panel of Figure~\ref{fig:blind_exp1}  shows the difference between the (smoothed) EMWA of the squared innovation after and before the additional sources have been introduced.  
The two bright hotspots at the injection locations are clearly visible. The center panel of Figure~\ref{fig:blind_exp1} shows the trajectory of $M_t$, which exceeds the threshold $k M_{\mathrm{base}}$ right after the injection of the additional two sources. In particular, an alarm is triggered at time step 467, yielding a detection delay of 17 steps ($3.8\%$) relative to the true injection at step 450.
The right panel adds the contours on a post-injection wave field snapshot. The spatial localization of the additional sources (i.e., anomalies) is highly accurate: the top two energy peaks are located at $(2.20, 7.00)$ and $(7.20, 1.50)$, matching the true injection locations to within $1\%$ of the domain size. 

\subsection{Experiment II: Heterogeneous Seismic Medium with Ricker Wavelets}

\begin{figure}[h!]
    \centering
    \includegraphics[width=0.85\linewidth]{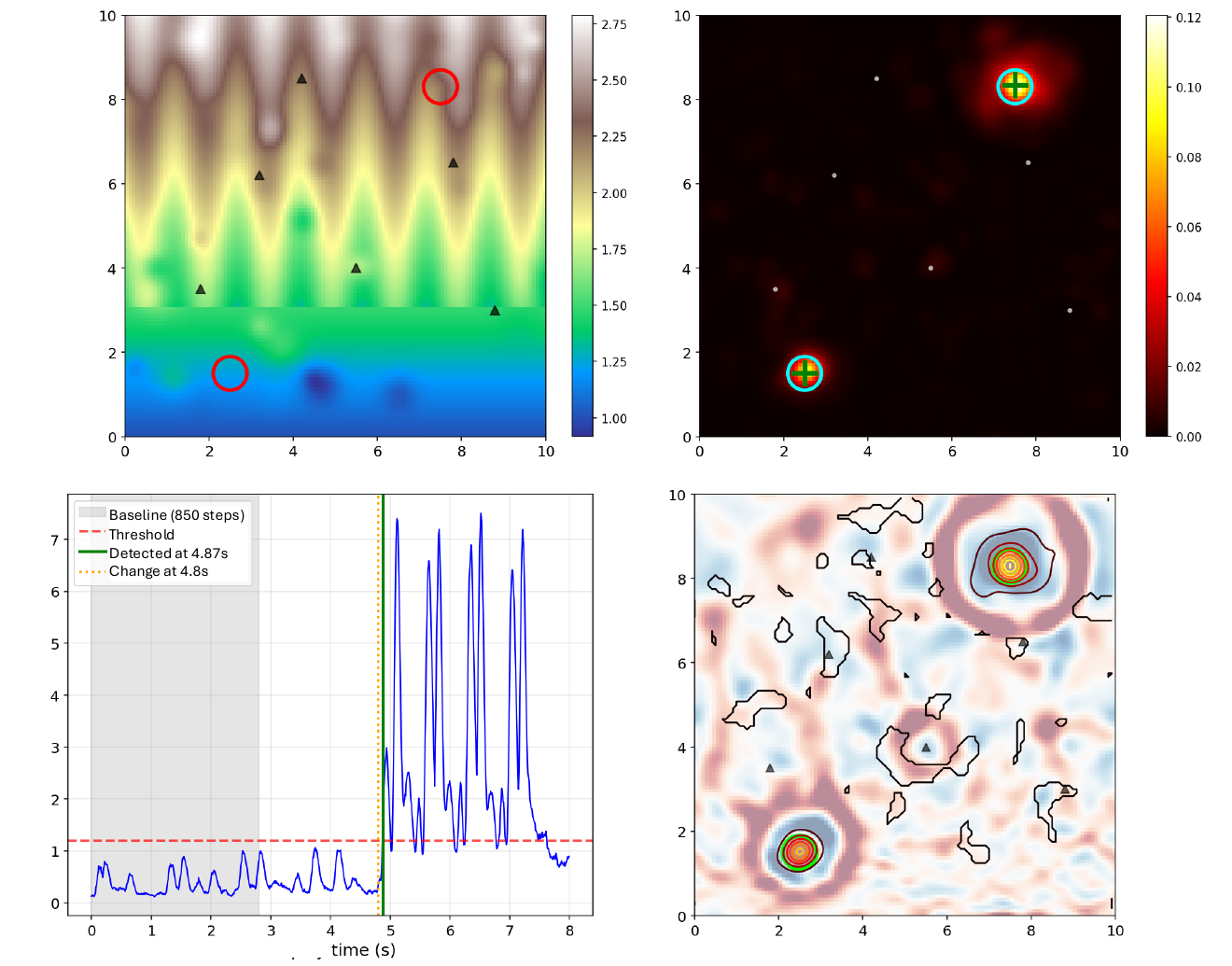}
    \caption{Heterogeneous velocity field $c(\bm{s})$ (top-left), EWMA of the squared innovation after and before the additional sources have been introduced (top-right), trajectory of $M_t$ (bottom-left), anomaly contours on a post-injection wave field snapshot (bottom-right).}
    \label{fig:blind_exp2}
\end{figure}

\textbf{Data and Experiment Setup.} Example II considers a more realistic example that involves seismic wave propagation in a heterogeneous medium. The same governing wave equation is considered as shown in (\ref{eq:wave_example1}). Instead of uniform wave field in Example I, we consider a heterogeneous velocity model inspired by the Marmousi benchmark \citep{versteeg1994marmousi} given as follows (also shown in the top-right panel of Figure \ref{fig:blind_exp2}):
\begin{equation}
c(\bm{s}) = 1.0 + 1.6\,\frac{s_2}{L} + 0.15\sin\!\Bigl(\frac{2\pi s_1}{1.5}\Bigr)
    \cdot\mathbf{1}_{s_2/L>0.3},
\end{equation}
where $\bm{s}=(s_1,s_2) \in [0,L]^2$, $s_1$ and $s_2$ denote the horizontal and vertical (depth) coordinates respectively, $L=10$ is the domain side length, and $s_2/L \in [0,1]$ is the normalized depth. The layered profile is perturbed by 40 random Gaussian scatterers, yielding a velocity field ranging from 0.7 to 3.2.
Six background seismic sources emit repeated Ricker wavelets (the standard seismological source-time function $R(t;f_0,t_0)=(1-2(\pi f_0(t-t_0))^2)\exp(-(\pi f_0(t-t_0))^2)$) with dominant frequencies $f_0\in[2.2,3.2]$\,Hz, amplitudes 40--60, and recurrence intervals of 1.2--1.5\,s. The simulation runs on a $120\times 120$ grid with $\Delta t=0.008$, damping $\gamma=0.18$, decay $\eta=0.008$, Gaussian observation noise with a standard deviation $0.04$, and process noise with a standard deviation $0.02$.
The model is solved for a total of 1500 time steps. Similar to Example I, the model parameters are estimated from MLE using data in the first 850 time steps. At time step 1100, two anomalous Ricker sources of amplitudes 300 and 280 are introduced at $(2.5,1.5)$ and $(7.5,8.3)$. 

\begin{figure}[h!]
    \centering
    \includegraphics[width=0.85\linewidth]{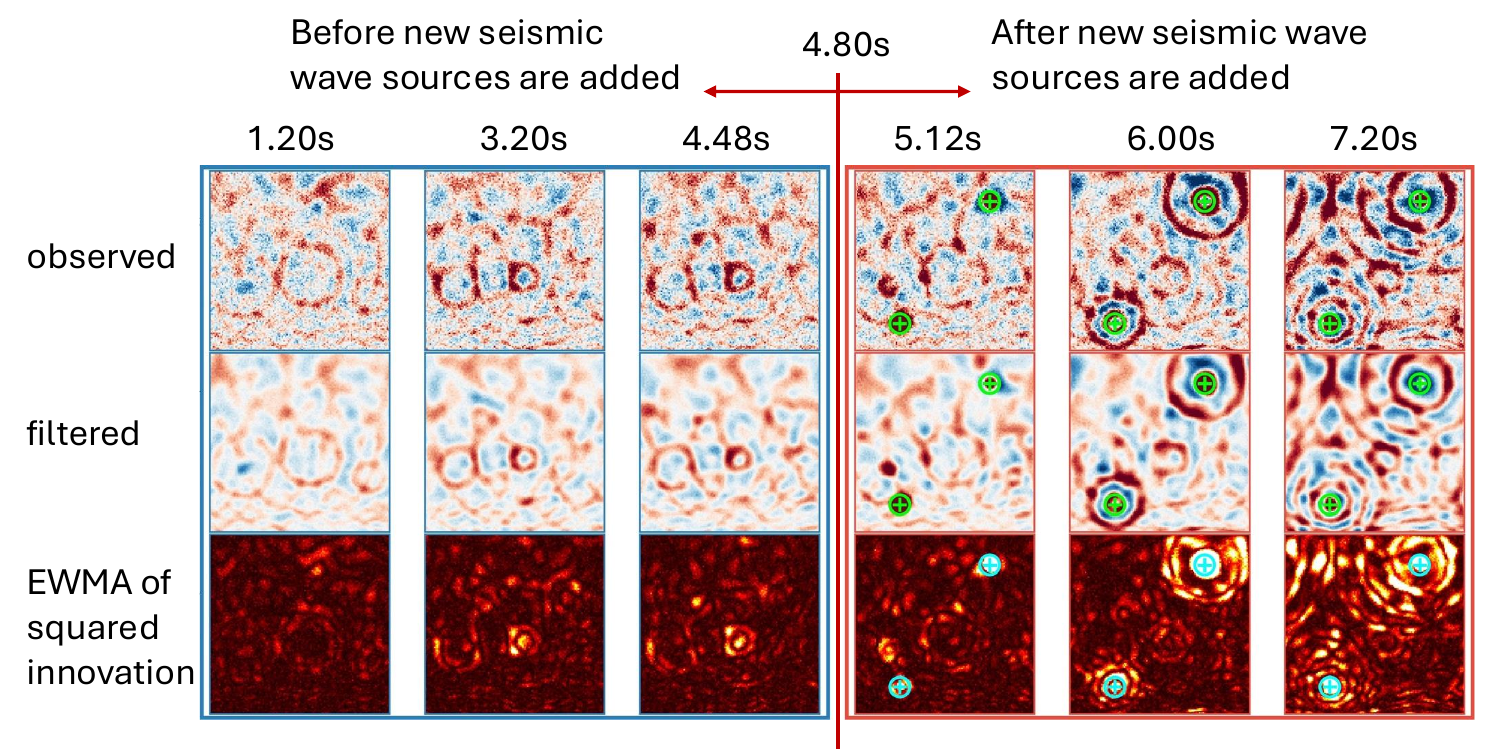}
    \caption{Observed (top) and filtered (middle) seismic wave propagation, and the EWMA of the squared innovation (bottom) with the circles indicating the locations where new seismic wave sources are introduced.}
    \label{fig:kf_panels_exp2}
\end{figure}
\textbf{Results.} 
Figure \ref{fig:kf_panels_exp2}  shows both the observed (top row) and filtered (middle row) processes at 6 chosen time steps: 1.2s, 3.2s, 4.48s, 5.12s, 6.00s and 7.10s. The filtered process (reconstructed through the inverse Fourier transform from the filtered $\bm{\theta}$ from (\ref{eq:solution})) well captures the wave propagation process. Similar to Example I, the one-step prediction error is calculated for each spatial location. 
These residuals are white noise with a mean of zero before the two anomalies are introduced at $t=4.80$s.
The bottom row of Figure \ref{fig:kf_panels_exp2} shows the EWMA of the squared innovation in this numerical example (using the same setting as in Example I). 
It is seen from this figure that the EWMA of the innovation energy remains spatially diffuse (reflecting the background seismicity). After $t=4.80$s, the EWMA map shows two persistent hotspots at the anomalous seismic wave source locations that intensify over time. 


The top-left panel of Figure~\ref{fig:blind_exp2} shows the heterogeneous velocity field $c(\bm{s})$, with the layered depth-dependent structure and sinusoidal lateral perturbations visible above $s_2/L=0.3$; triangles mark the six background sources and circles mark the two injected anomaly locations. Similar to Example~I, the top-right panel shows the difference between the (smoothed) EWMA of the squared innovation after and before the additional sources have been introduced. The two bright hotspots are clearly visible indicating the new seismic sources. The bottom-left panel of Figure~\ref{fig:blind_exp2} shows the trajectory of $M_t$, which exceeds the threshold $k M_{\mathrm{base}}$ right after the injection of the additional two sources. In particular, an alarm is triggered at step 608, yielding a detection delay of 8 steps ($1.3\%$) relative to the true injection at step 600. The detector produces zero false alarms throughout the 250-step monitoring gap between training and injection. The bottom-right panel shows the anomaly contour map overlaid on a post-injection wave field snapshot; the contours concentrate tightly around the two injected source locations, confirming accurate spatial localization. The spatial localization identifies peaks at $(7.50, 8.33)$ and $(2.50, 1.50)$, both within $0.5\%$ of the true locations.

\subsection{Example III: Advection-Diffusion Process} 

\textbf{Data and Experiment Setup.} In Example III, the proposed model is applied to model and detect a change point during a stochastic advection-diffusion process. The following stochastic PDE with a source term is used to generate the data:
\begin{equation} \label{eq:spde2}
\begin{aligned}
\dot{\xi}(t,\bm{s}) & = -\bm{\mu}^T \triangledown \xi(t,\bm{s}) +  \triangledown \cdot \bm{\Sigma} \triangledown \xi(t,\bm{s}) \\ & - \eta \xi(t,\bm{s}) + Q(t,\bm{s}) + \epsilon(t,\bm{s}),
\end{aligned}
\end{equation}
where $\bm{s} \in [0,1]^2$.
In this example, the SPDE (\ref{eq:spde2}) is solved on a discretized spatial and temporal domain. The spatial domain is discretized by a $40\times 40$ regular grid, and the temporal domain is discretized into equal-length intervals with a size of $\Delta_t=0.1$. Here, we let $T=3.3$, leading to 33 simulation time steps \textit{excluding} the simulation of the initial condition.  
\begin{figure}[h!] 
    \centering
\includegraphics[width=0.6\linewidth]{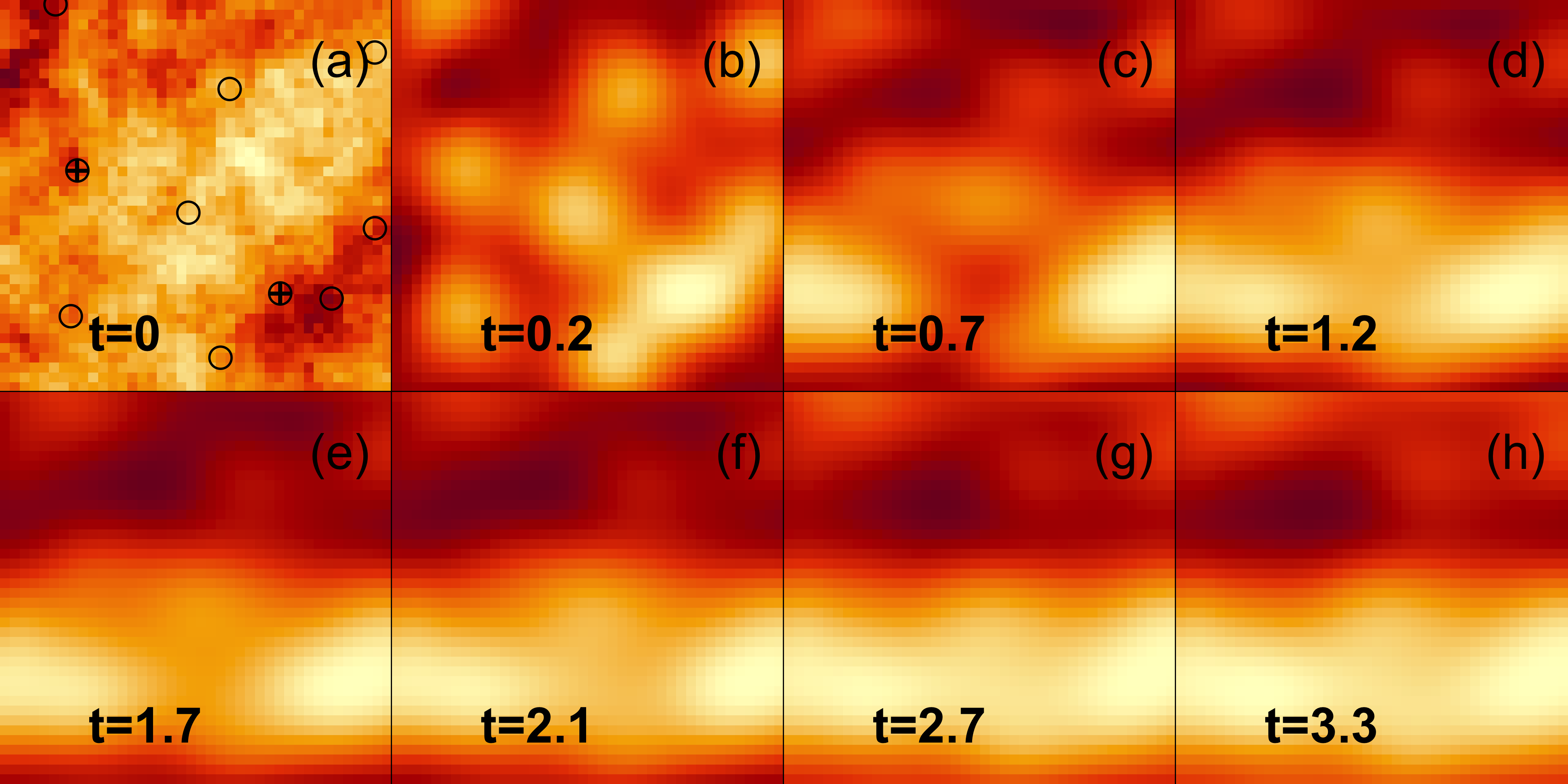}
\caption{A simulated advection-diffusion process with a change in the source term at $t=2$.} \label{fig:data1}
\end{figure}

The initial condition $\xi(0,\bm{s})$ is simulated from a Gaussian Random Field (GRF) with an Exponential covariance, which has a partial sill of 1 and range 0.25; see Figure \ref{fig:data1}(a). 
The advection, diffusion and decay are respectively set to $\bm{\mu}=(1,0)^T$, $\bm{\Sigma}=\text{diag}(0.02,0.02)$ and $\eta=0.1$. The white-in-time noise  $\epsilon(t,\bm{s})$ is a zero-mean GRF.  
The source term $Q(t,\bm{s})=\sum_{i=1}^{10}Q_i(t,\bm{s})$ involves 10 constant sources whose locations are also indicated in Figure \ref{fig:data1}(a). For each source function $Q_i(t,\bm{s})$, $i=1,\cdots,10$, we let $Q_i(t,\bm{s})= 10\tilde{f}^{-1}  \times f(\bm{s};\bm{\mu}^{Q}_i,\bm{\Sigma}^{Q}_i)$ where $f(\cdot)$ is the densify function of a bivariate Gaussian distribution, $\bm{\mu}^{Q}_i$ specifies the  location of the center of the source,  $\bm{\Sigma}^{Q}_i = \text{diag}(10^{-3},10^{-3})$, and $\rho$ is the value of the density function $f$ evaluated at its center $\bm{\mu}^{Q}_i$. 
A change point is introduced at $t=2$ (i.e., the 20th simulation time step). Starting from this change point, two of the ten sources become weaker. In particular, for the two sources terms indicated by ``$\oplus$'' in Figure \ref{fig:data1}(a), the source function becomes $Q_i(t,\bm{s})= 5\rho^{-1}\times f(\bm{s};\bm{\mu}^{Q}_i,\bm{\Sigma}^{Q}_i)$. 

Figure \ref{fig:data1}(a)-(h) show the simulated process at selected times. Starting from its initial condition, we see that the process gradually reaches a more stationary state. Although a change point is introduced to the source term $Q(t,\bm{s})$ at time $t=2$, \textit{such a subtle change can hardly be detected by the naked eye}.

\textbf{Results and Comparison.}
For comparison purposes, two models are applied to the simulated data to detect the change in the source term introduced at time $t=2$. 
The first model is the one proposed in this paper. Since (\ref{eq:spde2}) is a real-valued process, the model is described in Section \ref{sec:real_value}. 

The second model (referred to as the conventional approach) is the state-space model, described in \cite{liu2022statistical}, for advection-diffusion processes governed by the SPDE (\ref{eq:spde2}). A brief description of this competing approach is given as follows. 
Let $\bm{Y}(t)=(Y(\bm{s}_1,t),Y(\bm{s}_2,t),\cdots, Y(\bm{s}_n,t))$ be an $n$-dimensional vector representing the discretized values of (\ref{eq:spde2}) at a set of locations $\{\bm{s}_1, \bm{s}_2, \cdots, \bm{s}_n\}$ (in this example, $n=1600$). The advection-diffusion process (\ref{eq:spde2}), with a source term, can be represented by a dynamical system that
is linear in state
\begin{equation} \label{eq:model2}
\begin{aligned}
  & \begin{bmatrix}
 \bm{\alpha}(t_{k}) \\
  \bm{\beta}(t_{k}) \\
\end{bmatrix}=\begin{bmatrix}
\bm{G} & \bm{I} \\
\bm{0} & \bm{I} \\
\end{bmatrix}\begin{bmatrix}
 \bm{\alpha}(t_{k-1}) \\
  \bm{\beta}(t_{k-1}) \\
\end{bmatrix}+\bm{e}(t_{\bm{k}}) \\
  & \bm{Y}(\bm{s},t_{\bm{k}}) = \begin{bmatrix}
 \bm{\phi}& \bm{0} \\
\end{bmatrix}\begin{bmatrix}
 \bm{\alpha}(t_{k}) \\
  \bm{\beta}(t_{k}) \\
\end{bmatrix} + \bm{\varepsilon}(\bm{s},t_{k}). 
\end{aligned}
\end{equation}

Details of the model (\ref{eq:model2}) are available in \cite{liu2022statistical}. Here, $\bm{\phi}$ is an $n\times n$ matrix with its $i$-th row being $\bm{\phi}(\bm{s}_i) = (\phi_1(\bm{s}_i),...,\phi_n(\bm{s}_i) )$ where $\phi_j(\bm{s})=\exp(i\bm{k}_j^T\bm{s})$ is the Fourier basis function and $\bm{k}_j$ is the spatial wavenumber, $\bm{I}$ is the identity matrix, and  
$\bm{G}$ is an $n\times n$ state transition matrix operator depending on the parameters in (\ref{eq:spde2}),   $\bm{\varepsilon}$ and $\bm{e}$ are multivariate Gaussian noise vectors, and $\bm{\alpha}(t) \in \mathbb{R}^n$ and $\bm{\beta}(t) \in \mathbb{R}^n$ are the state vectors. 
For this conventional approach, it is important to note that $\bm{\beta}(t)$ captures the growth-decay of the process due to the source term $Q(t,\bm{s})$ and is often modeled as an AR(1) process. The inverse Fourier transform of $\bm{\beta}(t)$ gives the amount of growth, or decay, of the process over a fixed time interval over the spatial domain. 
Hence, to detect the change point associated with $Q(t,\bm{s})$ at time $t=2$ in this example, the filtered dynamics of $\bm{\beta}(t)$ plays a critical role if the conventional model (\ref{eq:model2}) is adopted.

\begin{figure}[h!] 
    \centering
\includegraphics[width=0.6\linewidth]{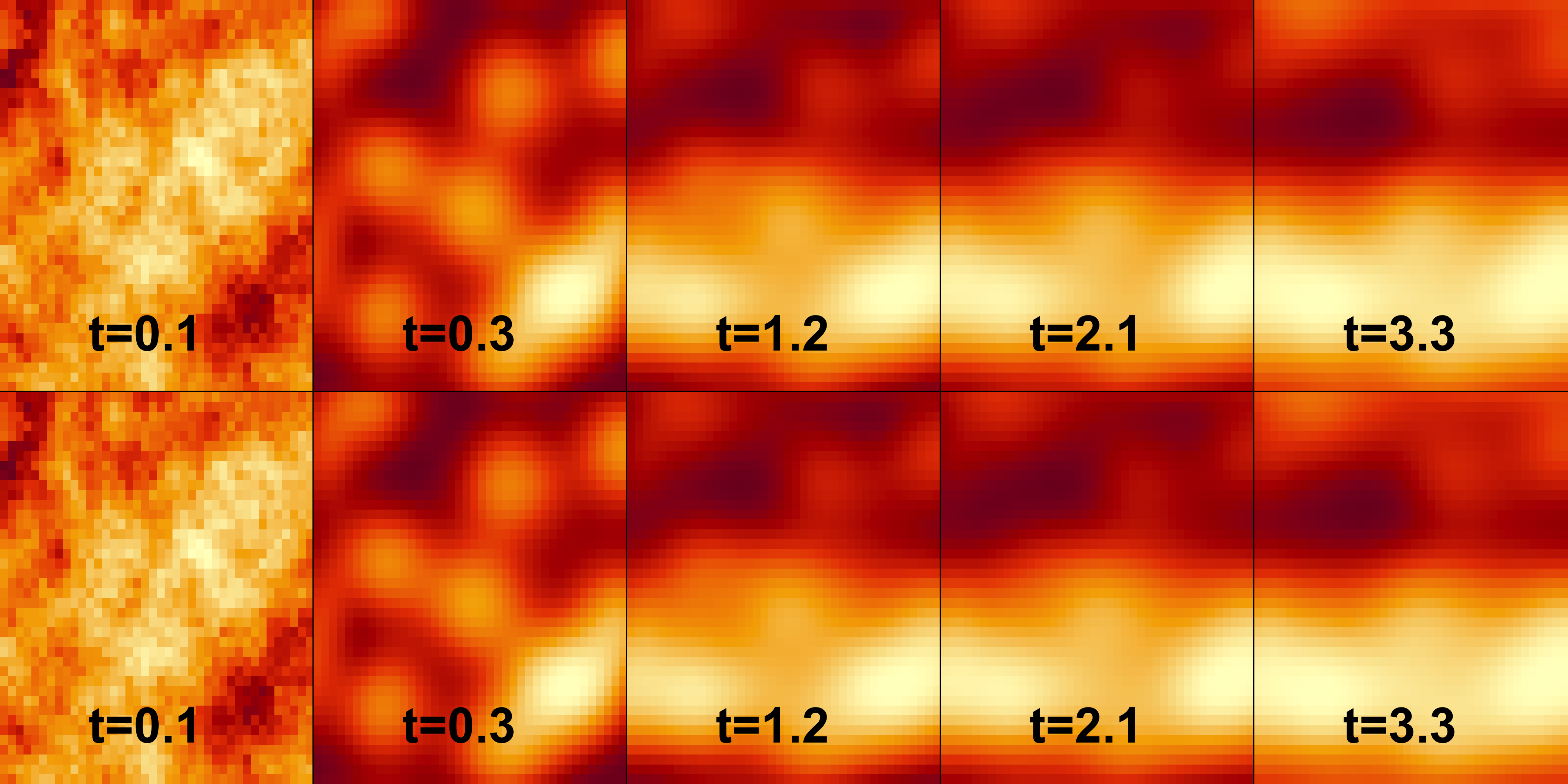}
\caption{Filtered processes from the proposed (above) and the conventional (below) models, and both models yield satisfactory performance.} \label{fig:filter1}
\end{figure}
The Kalman Filter is used to estimate the state dynamics for both the proposed and convectional linear state-space models. Note that, in this numerical example based on simulation data, our primary goal is to investigate if the filtered states can detect the change in the source term at time $t=2$, and all model parameters are given for both models for a fair comparison. 
Figure \ref{fig:filter1} firstly shows the re-constructed processes using the filtered system states at selected times. The row above shows the filtered process obtained from the proposed model, while the row below shows the process obtained from the convectional model (\ref{eq:model2}). By comparing Figures \ref{fig:data1} and \ref{fig:filter1}, it is seen that both models yield seemingly satisfactory performance in terms of obtaining the filtered process. 

\begin{figure}[h!] 
    \centering
\includegraphics[width=1\linewidth]{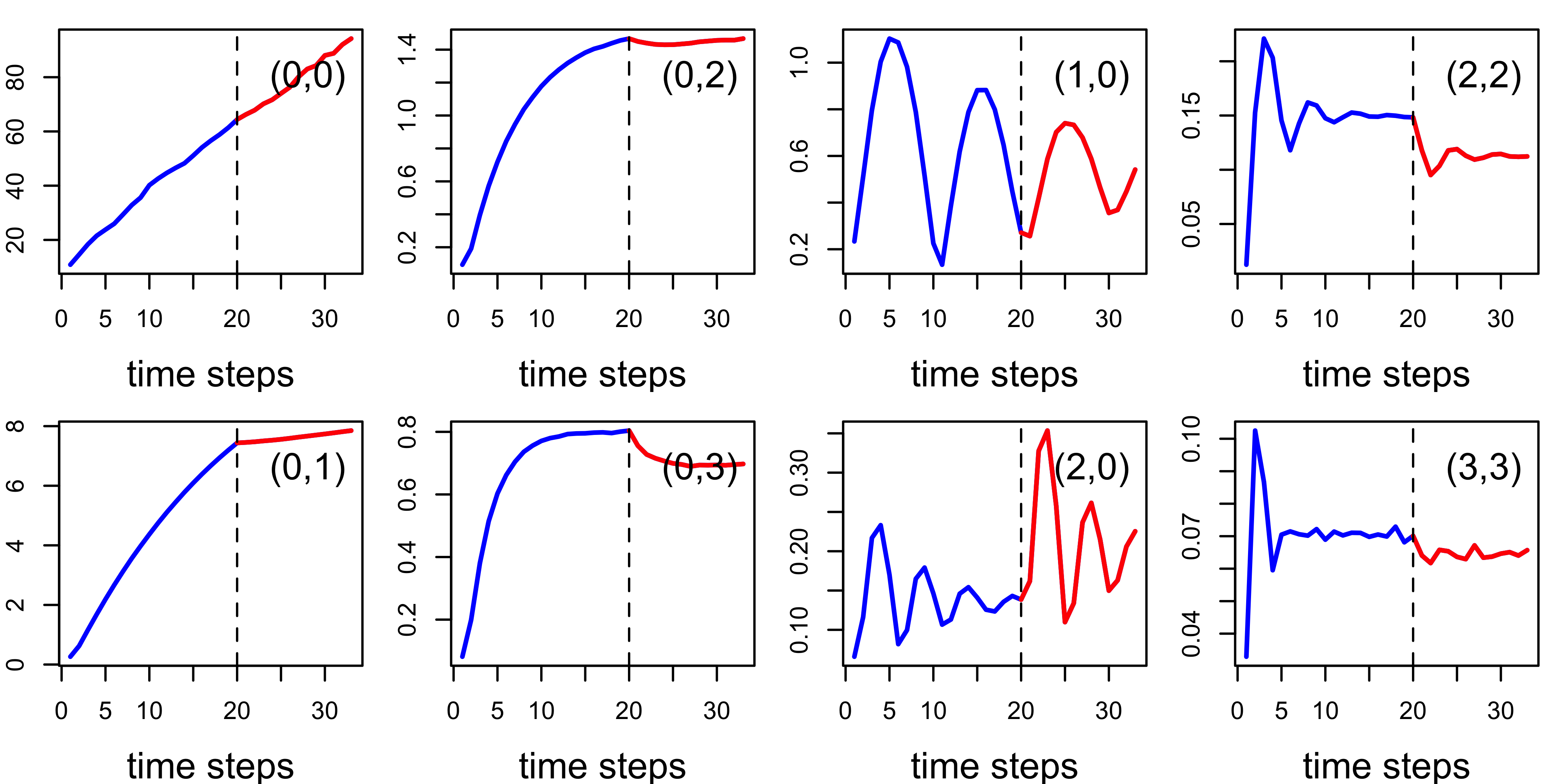}
\caption{Filtered $\alpha_{\bm{k}}(t)$ for selected spatial frequencies from the proposed approach.} \label{fig:filter_alpha}
\end{figure}
\begin{figure}[h!] 
    \centering
\includegraphics[width=1\linewidth]{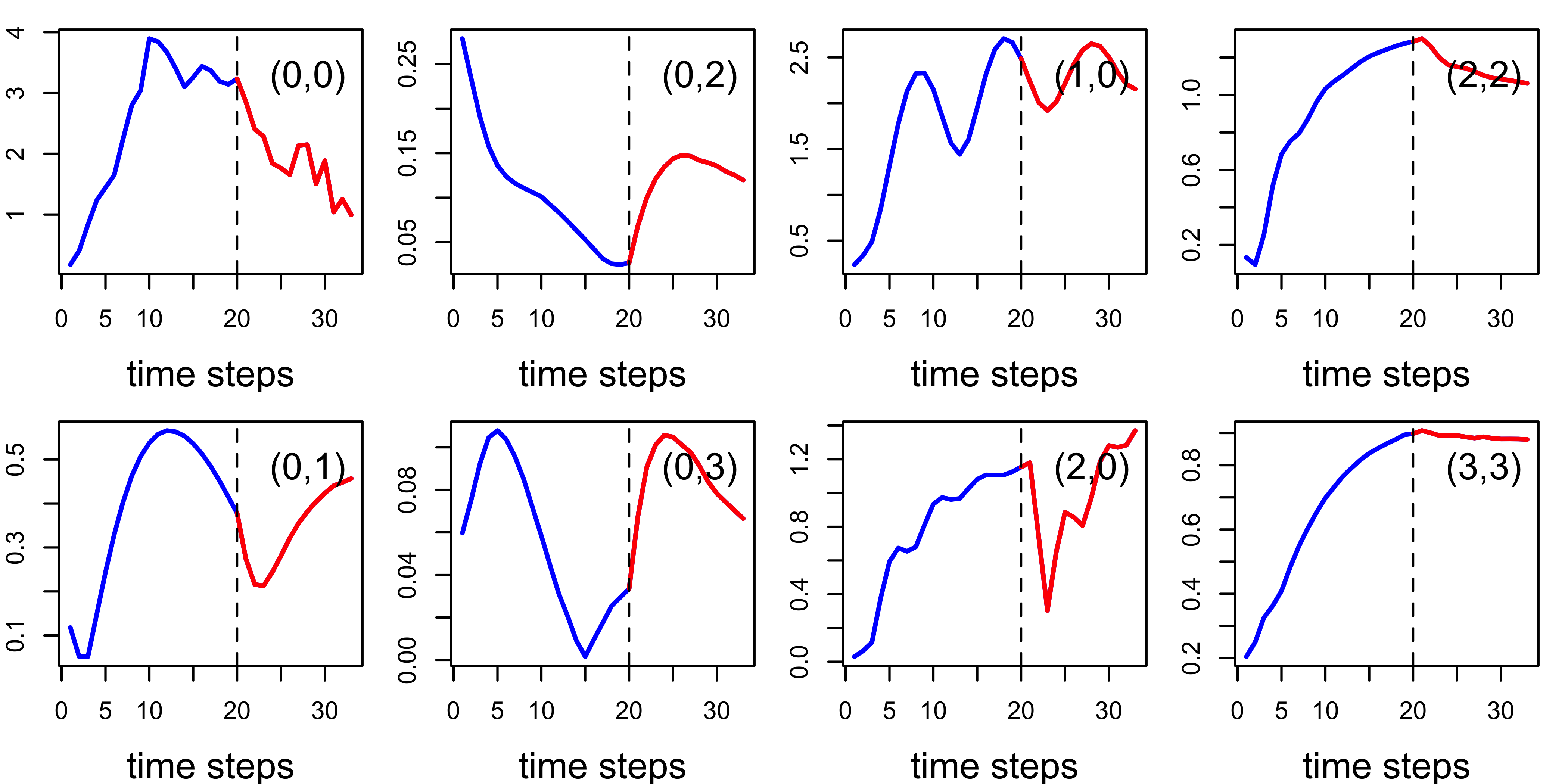}
\caption{Filtered $\beta_{\bm{k}}(t)$ for selected spatial frequencies from the proposed approach.} \label{fig:filter_beta}
\end{figure}
Next, we investigate which model performs better in detecting the change of the source term at time $t=2$ (i.e., at the simulation step 20). 

$\bullet$ For the proposed model, Figures \ref{fig:filter_alpha} and \ref{fig:filter_beta} respectively show the filtered $\alpha_{\bm{k}}(t) = \sqrt{(\alpha_{\bm{k}}^R(t))^2 + (\alpha_{\bm{k}}^I(t))^2}$ and $\beta_{\bm{k}}(t) = \sqrt{(\beta_{\bm{k}}^R(t))^2 + (\beta_{\bm{k}}^I(t))^2}$ for a set of selected wavenumbers from the set, $\{(0,0), (0,1), (0,2), (0,3), (1,0), (2,0),(3,0), (1,1), (2,2), (3,3)\}$. These wavenumbers are selected from the low-frequency region of the spectral space and are associated with the most important dominate pattern of the process (of course, it is possible to show the graph for other wavenumbers).  
Also recall that, $\alpha^R_{\bm{k}} = \dot{\beta}^R_{\bm{k}}$ and $\alpha^I_{\bm{k}} = \dot{\beta}^I_{\bm{k}}$ in the proposed model. Hence, $\alpha_{\bm{k}}(t)$ and $\beta_{\bm{k}}(t)$ are respectively associated with the process and its first-order derivative.

Some important observations are obtained from Figures \ref{fig:filter_alpha} and \ref{fig:filter_beta}:

$\diamond$ It is \textit{ineffective}, if not impossible, to detect the change of the source term by simply looking at the trend of the estimated overall mean (i.e., at the wavenumber $\bm{k}=(0,0)$). For $\bm{k}=(0,0)$, the coefficient $\alpha_{\bm{k}=(0,0)}(t)$ is exactly the filtered overall spatial mean of the process following the real-valued Fourier transform (the top left panel of Figure \ref{fig:filter_alpha}).  
Because the introduced change of the source term is relatively small and the data contain noise, the change of estimated overall spatial mean, before and after time step 20, does not provide a strong indicator of the change of source.   

$\diamond$ It is much more \textit{effective} to detect the change of the source term, at time step 20, by looking at the estimated trend of $\alpha_{\bm{k}}$ corresponding to other spatial frequencies; for example, (0,1), (0,2) and (0,3). The \textit{abrupt}, \textit{non-smooth and visually noticeable} change of the trend $\alpha_{\bm{k}}(t)$ at time step 20 indicates the change; see Figure \ref{fig:filter_alpha}.  

$\diamond$ The change of source at time step 20 is clearly noticeable from the estimated $\beta_{\bm{k}}(t)$ (which is related to the first-order derivative of the process) at multiple spatial frequencies, such as (0,1), (0,2), (0,3), (2,0), (2,2) and (3,3), as shown in Figure \ref{fig:filter_beta}.

\begin{figure}[h!] 
    \centering
\includegraphics[width=1\linewidth]{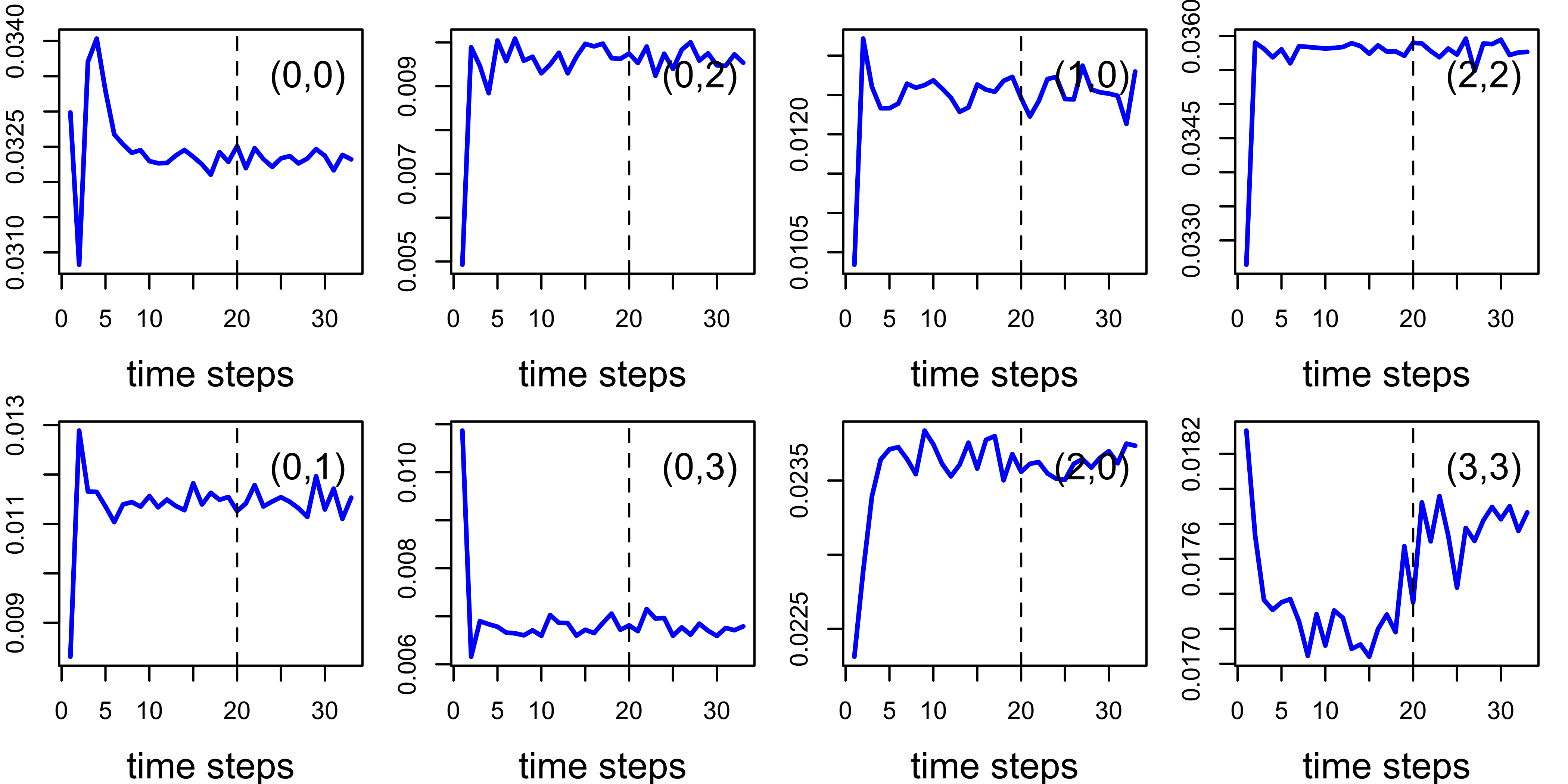}
\caption{Filtered $\alpha_{\bm{k}}(t)$ for selected spatial frequencies from the existing approach.} \label{fig:filter1_alpha}
\end{figure}
\begin{figure}[h!] 
    \centering
\includegraphics[width=1\linewidth]{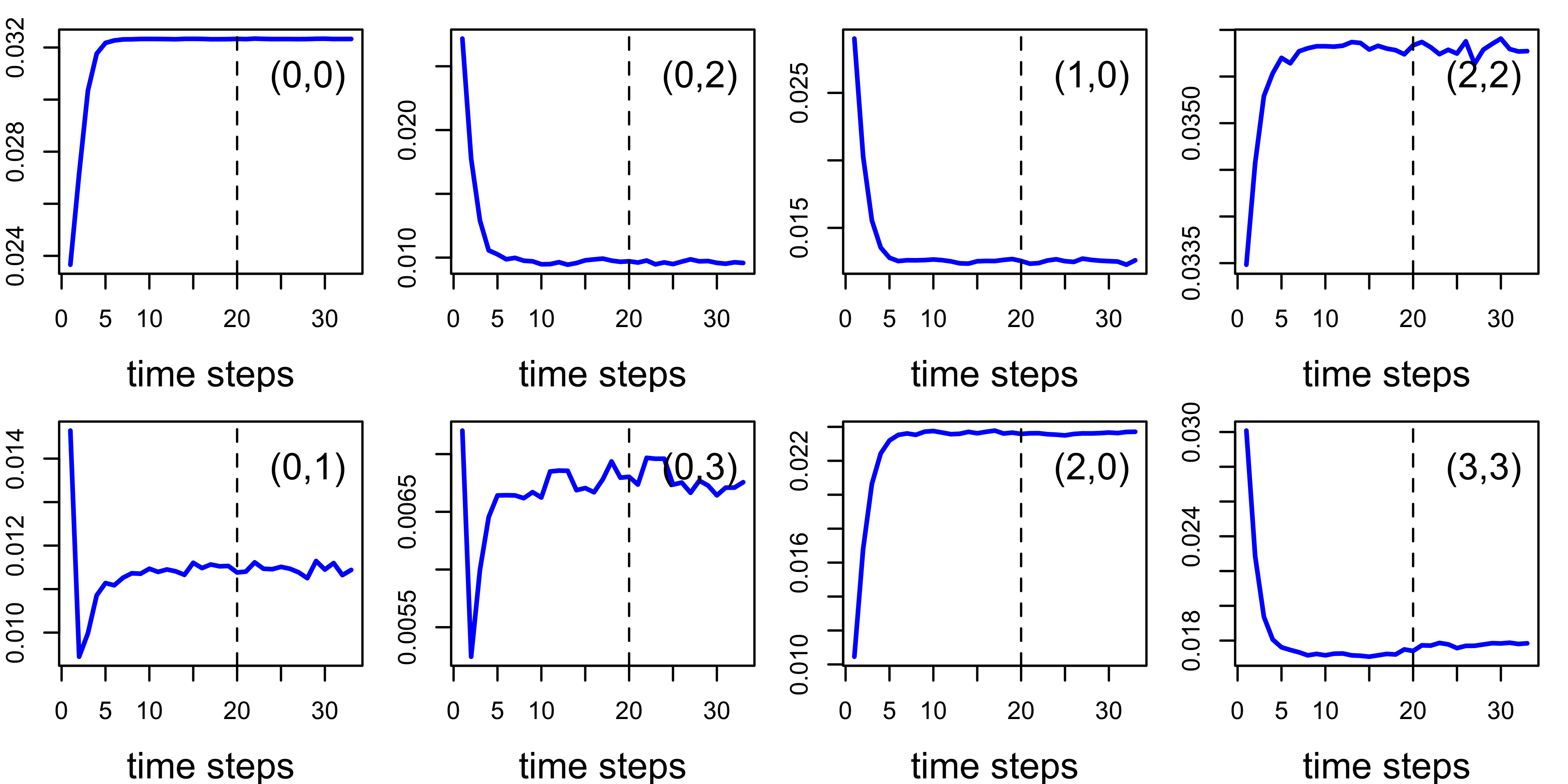}
\caption{Filtered $\beta_{\bm{k}}(t)$ for selected spatial frequencies from the existing approach.} \label{fig:filter1_beta}
\end{figure}

For comparison purposes, Figures \ref{fig:filter1_alpha} and \ref{fig:filter1_beta} show the filtered $\alpha_{\bm{k}}(t) = \sqrt{(\alpha_{\bm{k}}^R(t))^2 + (\alpha_{\bm{k}}^I(t))^2}$ and $\beta_{\bm{k}}(t) = \sqrt{(\beta_{\bm{k}}^R(t))^2 + (\beta_{\bm{k}}^I(t))^2}$, respectively, obtained from the conventional model (\ref{eq:model2}) for the same set of selected wavenumbers. However, neither the filtered dynamics of $\alpha_{\bm{k}}(t)$ nor $\beta_{\bm{k}}$ shows any clear indication of the change of the source term at time $t=2$.

In summary, this simulation-based comparison provides a strong evidence that the proposed approach yields some advantages over the existing approach in terms of detecting the first-order time derivative of the convolution-generated space-time processes (i.e., the change of the source term at $t=2$). The main reason is that the proposed state-space model in (\ref{eq:solution}) involves $\beta_1, \beta_2, \cdots, \beta_J$ as the second half of the state vector, and these coefficients are directly related to the first-order time derivative of the process as $\dot{x}(\bm{s},t)=\sum_{j=1}^{J}\phi_j(\bm{s})\beta_j(t)\equiv \bm{\phi}^T(\bm{s})\bm{\beta}(t)$. 

\subsection{Application: Remote-Sensing Wildfire Aerosol Process} 

\begin{figure}[h!]
    \centering
    \includegraphics[width=0.7\linewidth]{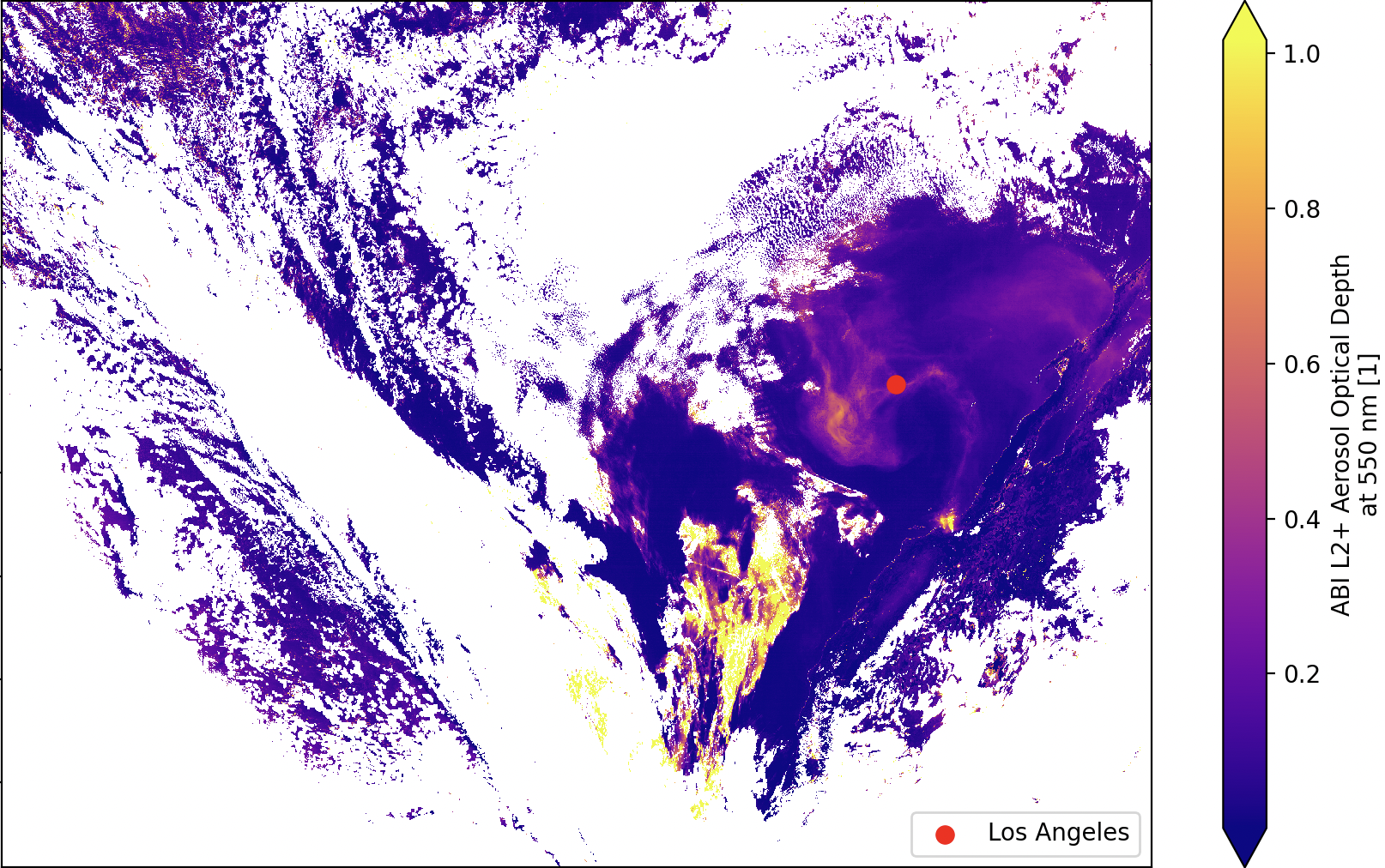}
    \caption{GOES-18 ABI Level-2 AOD distribution on January 9, 2025, near Los Angeles.}
    \label{fig:data2_full}
\end{figure}

The last numerical example involves the application of the proposed model to the remote-sensing data of wildfire smoke propagation. In recent years, wildfires have become more frequent, intense, widespread, and destructive globally. During wildfire events, aerosols directly impact public health, cloud radiative forcing, and solar energy production as high aerosol concentrations can drastically reduce photovoltaic efficiency. 
Aerosol Optical Depth (AOD) is a critical atmospheric indicator, reflecting the columnar loading of aerosols such as dust, smoke, or pollution \citep{Remer2005,Holben1998,IPCC2021}. As an illustrative example, Figure~\ref{fig:data2_full}
shows a snapshot of the spatial distribution of AOD during the Kenneth fire episode near Los Angeles, obtained from the GOES-18 satellite, on Jan 9, 2025. Normally, rapid AOD changes signify the onset and sudden growth of smoke plumes or fire intensity. For instance,  the satellite-based retrievals recorded dense AOD plumes around \emph{5{:}30\,PM} PST as the Kenneth fire rapidly grew fanned by strong Santa Ana winds. The increasing build-up AOD, or smoke, poses immediate threats to air quality, renewable energy generation, and yields critical information about the hidden fire dynamics which can hardly be directly monitored using optical measurements in the presence of dense smoke.  
first-order time derivative of AOD processes) and provide early indicators of abrupt events.

In this application example, we apply the proposed approach to model the remote-sensing AOD process and detect spatial areas with rapid AOD changes during the 2025 Los Angeles Kenneth Fire episode. The fire occurred in Jan 2025 in the West Hills area of Los Angeles County, California. 

\begin{figure}[h!] 
    \centering
    \includegraphics[width=0.8\linewidth]{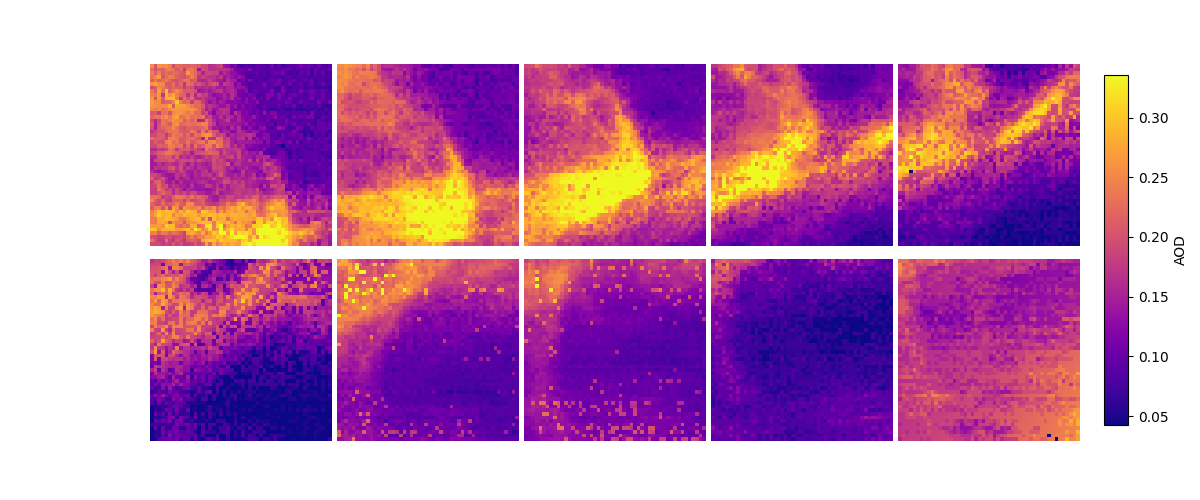}
    \vspace{-18pt}
    \caption{Ten evenly spaced (every 20 minutes) snapshots of AOD data streams from the cropped 40×40 window centered on Los Angeles.}
    \label{fig:data2_crop}
\end{figure}
\textbf{Data.} 
The 2025 Los Angeles wildfires AOD data are obtained from the GOES-18 ABI Level-2 Aerosol Optical Depth product, which retrieves columnar AOD at 550 nm on a 2 km geostationary grid every 10 minutes. 
Files are stored as NetCDF-4 in the public Google Cloud bucket  
\texttt{gs://gcp-public-data-goes-18/ABI-L2-AODC/2025/009/}.  
Each file contains variables \texttt{AOD}, \texttt{x}, \texttt{y}, and the \texttt{goes\_imager\_projection} metadata are used to map pixel indices to geographic coordinates. An example snapshot of AOD data is provided in Figure \ref{fig:data2_full}.  

We load each NetCDF file with \texttt{xarray} and read the projection parameters  from \\ \texttt{goes\_imager\_projection}.  Using \texttt{pyproj}, the native GOES grid is re-projected to longitude–latitude. We then locate the grid cell nearest to Los Angeles (34.0522° N, 118.2437° W), and extract a 40×40 pixel window centered on that cell.  By processing every third file (20 minute intervals) over a 10 hour span, we assemble a 40×40×30 array of AOD snapshots for this numerical example. Figure \ref{fig:data2_crop} presents ten evenly spaced snapshots from the cropped 40×40 window.  The top row illustrates the rapid buildup of AOD as the wildfire plume intensifies, while the bottom row shows the subsequent dispersion and gradual return toward ambient aerosol levels.

\begin{figure}[h!] 
    \centering
    \includegraphics[width=0.7\linewidth]{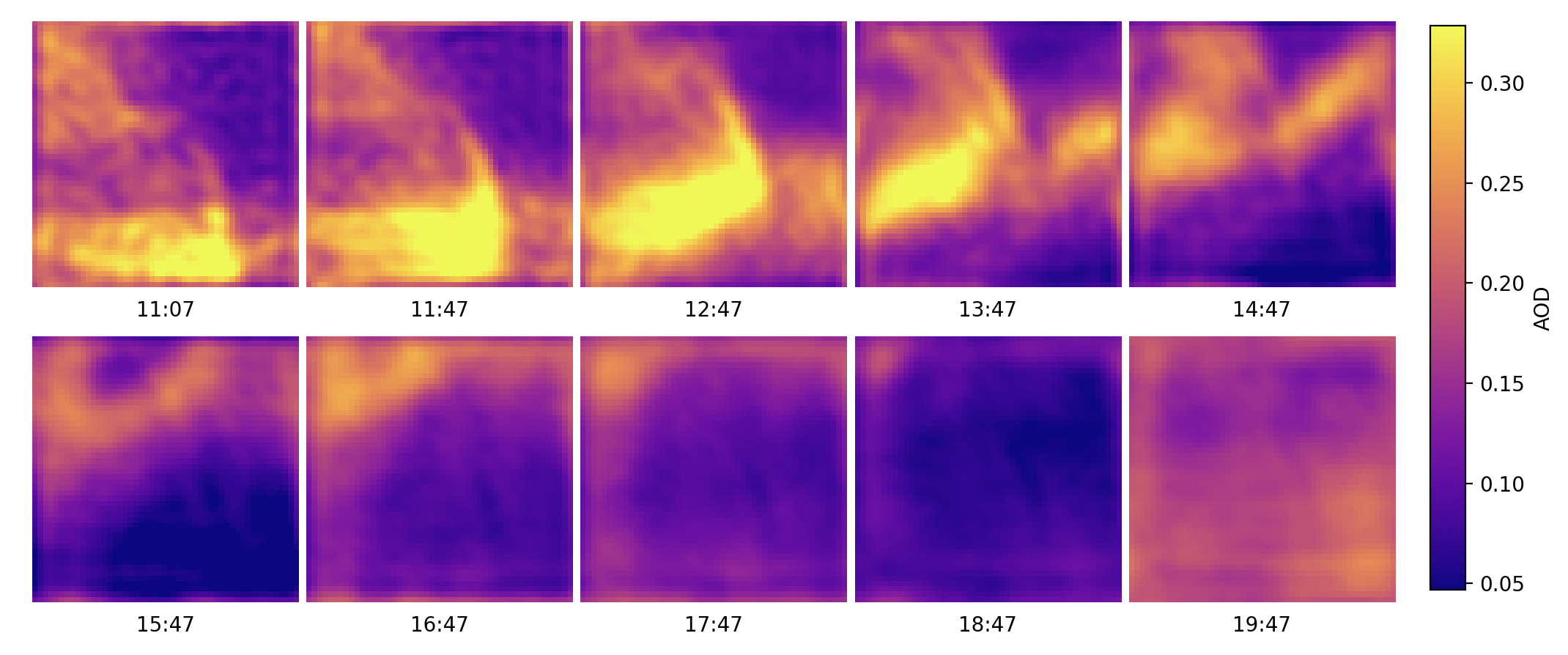}
    \vspace{-8pt}
    \caption{Filtered processes from the proposed model}
    \label{fig:data2_model1}
\end{figure}

\begin{figure}[h!] 
    \centering
    \includegraphics[width=0.7\linewidth]{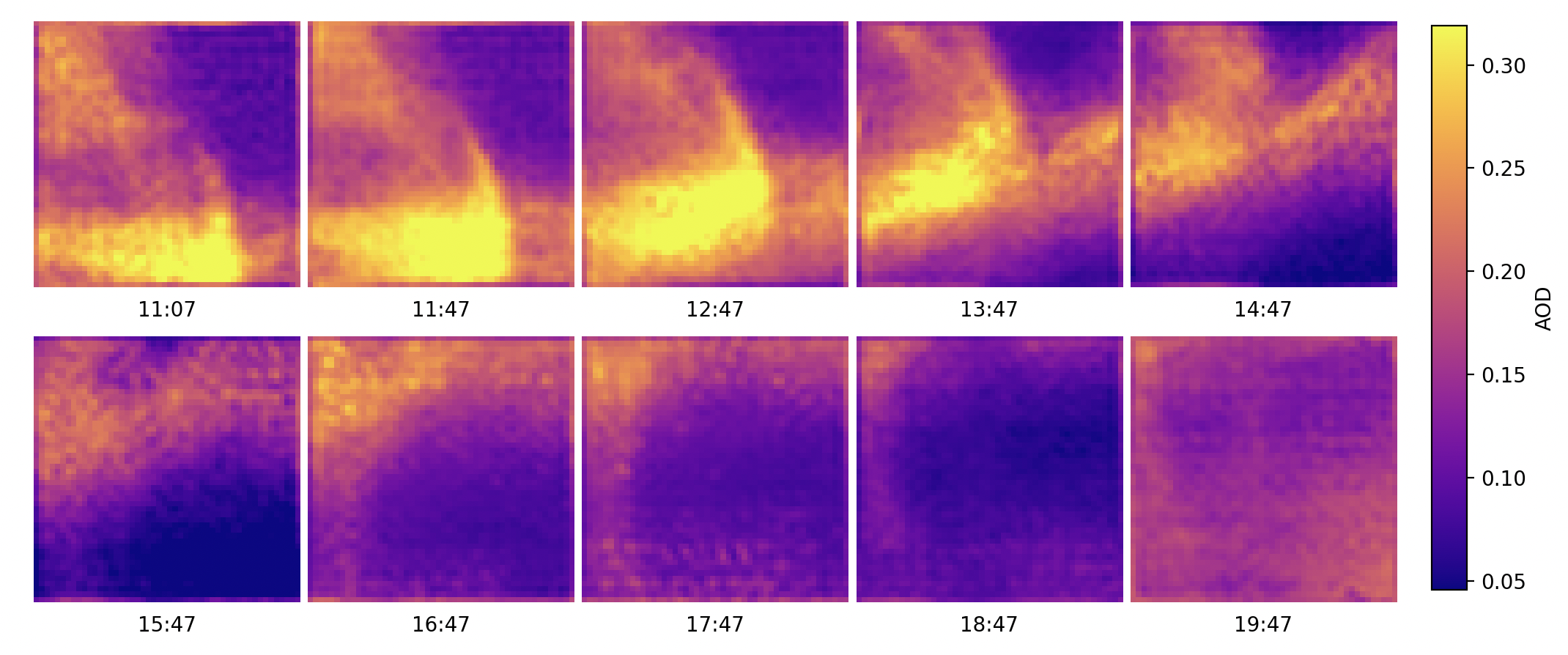}
    \vspace{-8pt}
    \caption{Filtered processes from the conventional models}
    \label{fig:data2_model2}
\end{figure}

\textbf{Results and Comparison.}
Similar to Example-III, we apply both the proposed and conventional models (\ref{eq:model2}) to the AOD data stream. Using the Adam optimizer with learning rate of $10^{-3}$, the maximum likelihood estimates of the unknown parameters are firstly obtained before the Kalman Filter is applied.
Figures \ref{fig:data2_model1} and \ref{fig:data2_model2} firstly present the filtered AOD process using both the proposed and the conventional models, respectively. In particular, for the proposed model, Figure \ref{fig:data2_model1} shows $\bm{\phi}^T(\bm{s})\hat{\bm{\alpha}}(t)$ using the filtered $\hat{\bm{\alpha}}(t)$. For the conventional model (\ref{eq:model2}), Figure \ref{fig:data2_model2} shows $\bm{F}^T(\bm{s})\hat{\bm{\alpha}}(t)$ using the filtered $\hat{\bm{\alpha}}(t)$. Both models seem to be capable of estimating the filtered AOD process. 
\begin{figure}[h!] 
    \centering
    \includegraphics[width=0.7\linewidth]{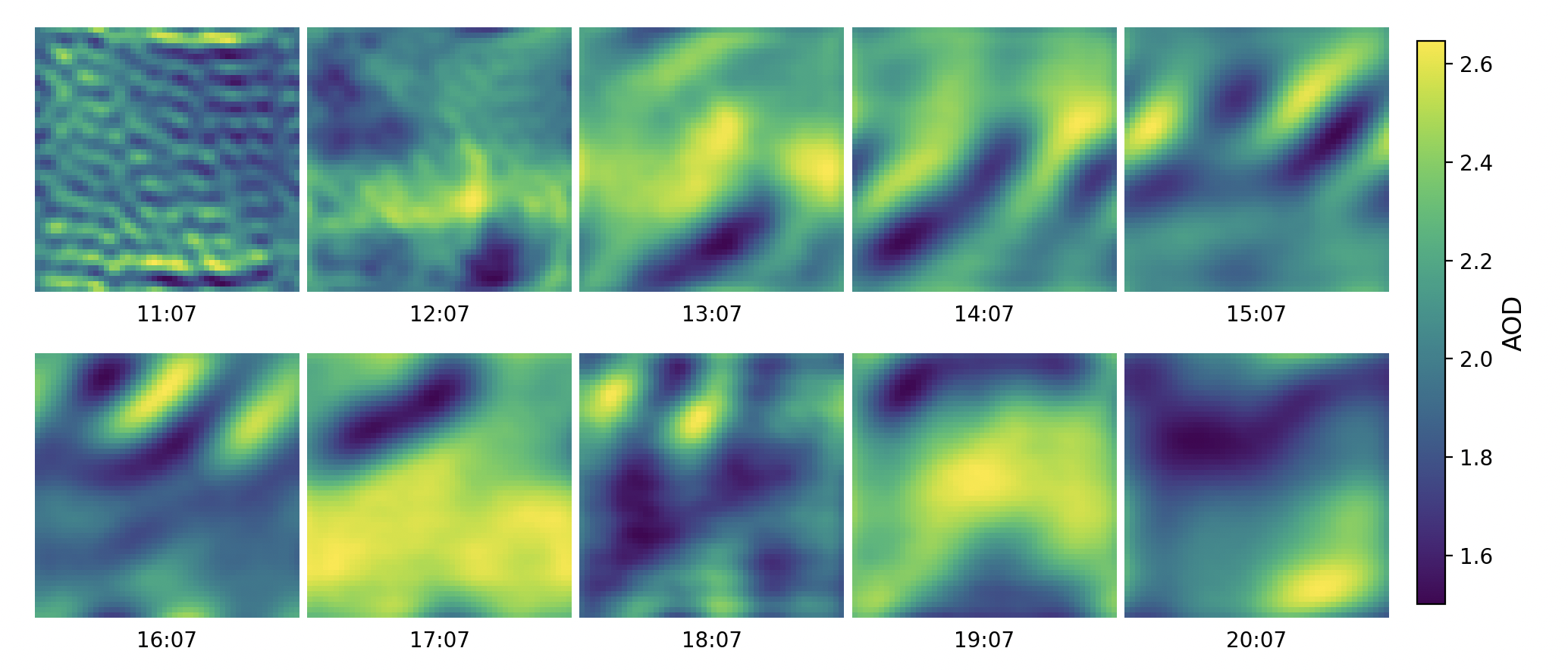}
    \vspace{-8pt}
    \caption{Estimated first-order time derivative of the process from the proposed model}
    \label{fig:data2_model1_beta}
\end{figure}
\begin{figure}[h!] 
    \centering
    \includegraphics[width=0.7\linewidth]{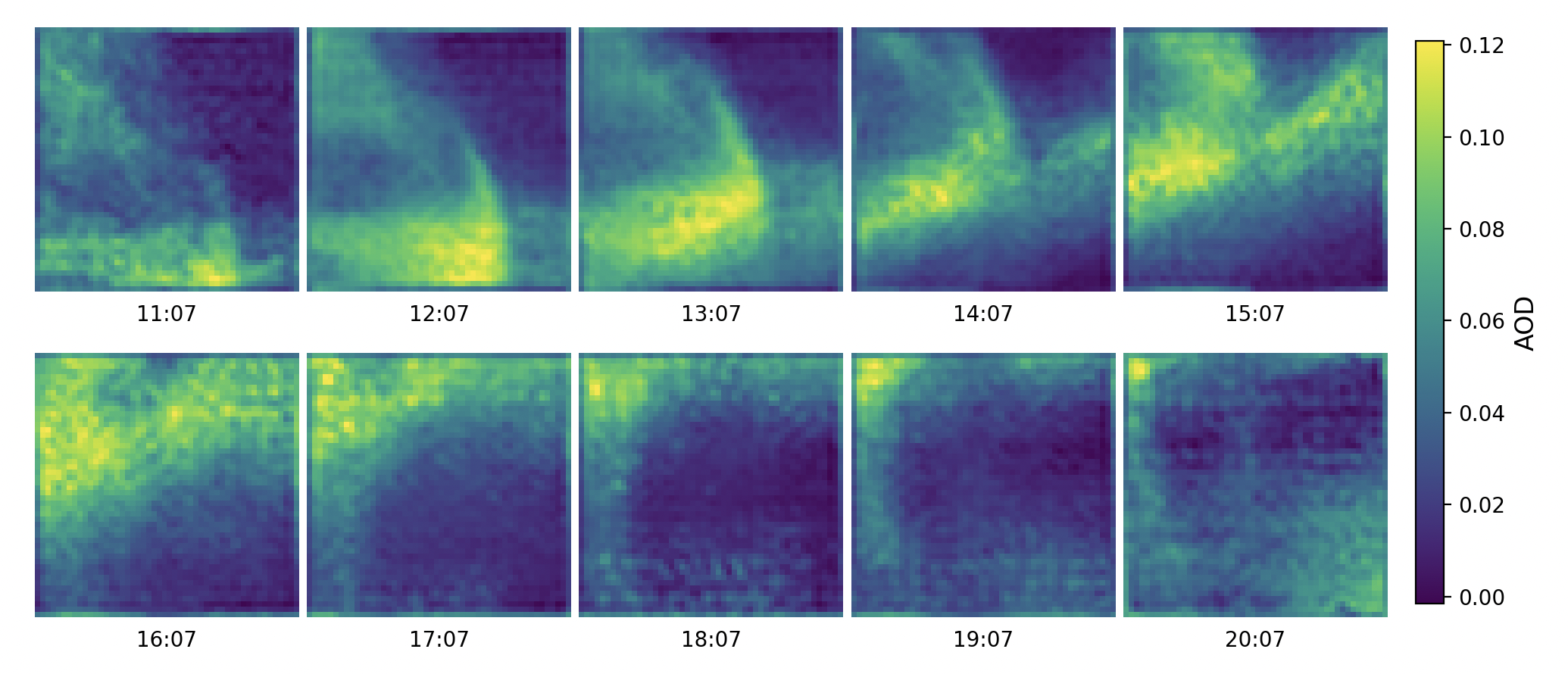}
    \vspace{-8pt}
    \caption{Estimated growth-decay from the conventional model}
\label{fig:data2_model2_beta}
\end{figure}

Next, we show the estimated first-order time derivative (using the proposed model) and the estimated growth-decay (using the conventional model). In particular, for the proposed model, $\bm{\phi}^T(\bm{s})\hat{\bm{\beta}}(t)$ gives the filtered first-order time derivative shown in Figure \ref{fig:data2_model1_beta}. For the proposed model, $\bm{F}^T(\bm{s})\hat{\bm{\beta}}(t)$ gives the estimated process growth-decay using the filtered $\hat{\bm{\beta}}(t)$ shown in Figure \ref{fig:data2_model2_beta}. 

A clear difference is observed between the results shown in Figures \ref{fig:data2_model1_beta} and \ref{fig:data2_model2_beta}. In Figure \ref{fig:data2_model1_beta}, the estimated first-order time derivative largely follows the plume's leading edge (the region with the most rapid change). This shows that the estimated transition matrix $\hat{\bm{G}}$ of the proposed model (\ref{eq:solution}) effectively captures the advection–diffusion dynamics of the AOD process. In contrast, the estimated growth-decay, shown in Figure \ref{fig:data2_model2_beta}, obtained from the conventional model 
apparently re-produces a similar pattern almost identical to the filtered process shown in Figure \ref{fig:data2_model1}. This observation is problematic and implies that the estimated transition matrix $\hat{\bm{G}}$ in the conventional model  (\ref{eq:solution}) \textit{cannot fully capture the dynamics of the AOD process}, and as a result, the estimated growth-decay $\bm{F}^T(\bm{s})\hat{\bm{\beta}}(t)$ serves as a bias correction, or compensation, to the filtered process $\bm{F}^T(\bm{s})\hat{\bm{\alpha}}(t)$ (rather than estimating the actural growth or decay of the process). \textit{This is a critical observation suggesting that the proposed model provides a stronger modeling capability over the existing approach}.



\begin{figure}[h!]
  \centering
  \includegraphics[width=\linewidth]{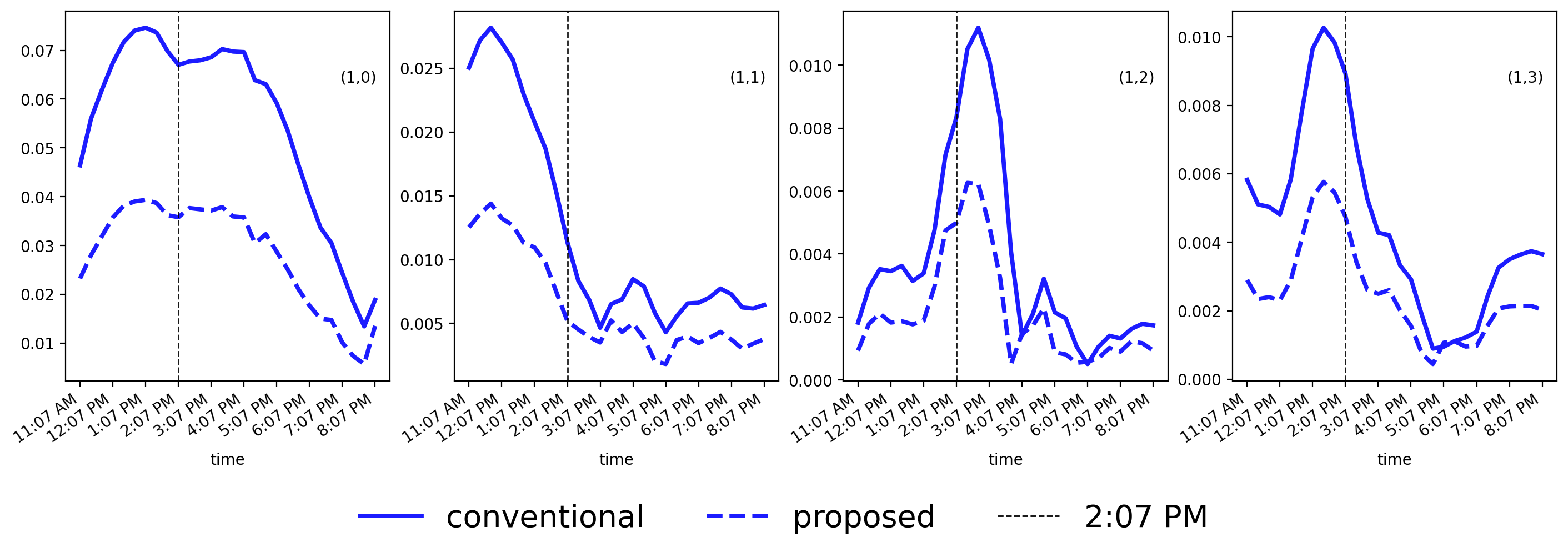}
  \vspace{-16pt}
  \caption{Comparison of $\alpha_{\bm{k}}(t)$ across some selected wavenumbers, $(1,0),(1,1),(1,2),(1,3)$. The vertical dashed line marks \emph{2:07\,PM} which indicates the time when the wildfire started.}
  \label{fig:freq_alpha_10to13}
\end{figure}

\begin{figure}[h!]
  \centering
  \includegraphics[width=0.6\linewidth]{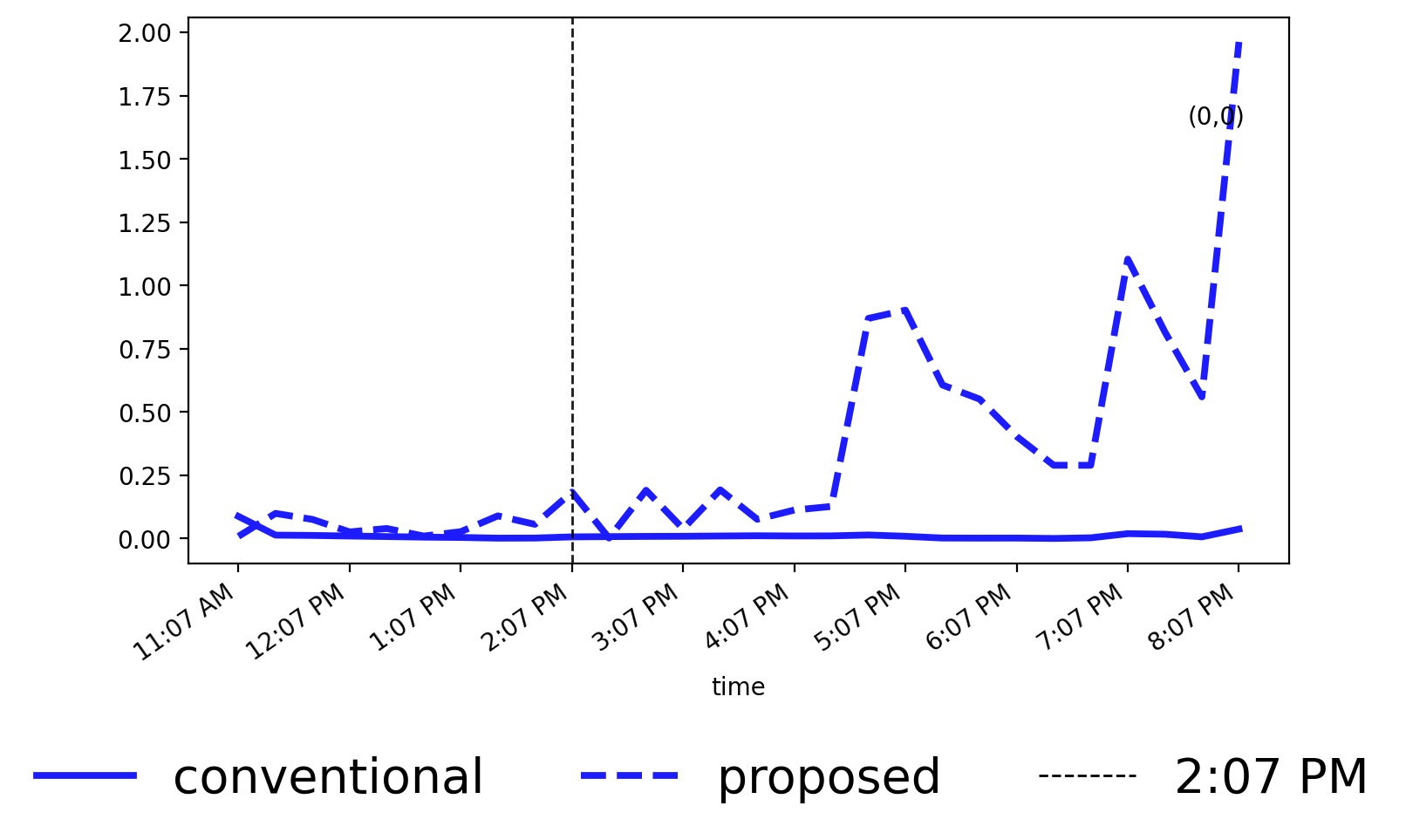}
  \vspace{-10pt}
  \caption{Comparison of $\beta_{\bm{k}}(t)$ at the selected wavenumber, $(0,0)$. The vertical dashed line (\emph{2:07\,PM}) indicates the onset of the Kenneth fire which grew rapidly by \emph{5{:}30\,PM} PST fanned by strong Santa Ana winds}
  \label{fig:freq_beta_00}
\end{figure}

When examining the filtered modal amplitudes $\alpha_{\bm{k}}(t) = \sqrt{(\alpha_{\bm{k}}^R(t))^2 + (\alpha_{\bm{k}}^I(t))^2}$ across some selected wavenumbers, $(1,0),(1,1),(1,2),(1,3)$, from the low-frequency region, both the proposed and conventional models trace highly similar trajectories for broad, slowly varying patterns as shown in Figure \ref{fig:freq_alpha_10to13}. The vertical dash line marks \emph{2:07\,PM} when the fire was initially reported. 
On the other hand, an interesting observation is obtained when examining the filtered $\beta_{\bm{k}}(t) = \sqrt{(\beta_{\bm{k}}^R(t))^2 + (\beta_{\bm{k}}^I(t))^2}$. In the proposed model, $\beta_{\bm{k}}(t)$ determines the time derivative of the AOD process, while in the conventional model, 
 $\beta_{\bm{k}}(t)$ is used to capture the growth or decay of the AOD process. Figure \ref{fig:freq_beta_00} shows the filtered $\beta_{\bm{k}}(t)$ at the lowest spatial frequency $(0,0)$ for both approaches. Around \emph{2:07\,PM}, the filtered $\beta_{\bm{k}}(t)$ from the proposed model starts to fluctuate and grow, which approximately coincides with the reported onset of the Kenneth fire near the border of Los Angeles and Ventura counties. More interestingly, the filtered $\beta_{\bm{k}}(t)$ from the proposed model shows a clear late afternoon surge (after \emph{4{:}00\,PM}) in line with public reports that the Kenneth Fire grew rapidly by \emph{5{:}30\,PM} PST fanned by strong Santa Ana winds. The conventional model, on the other hand, obviously fails to capture the sudden growth of AOD. The reason behind such an observation has already been provided in the discussions of Figures \ref{fig:data2_model1_beta} and \ref{fig:data2_model2_beta}. As discussed above, the estimated transition matrix $\hat{\bm{G}}$ in the conventional model  (\ref{eq:solution}) does not fully capture the dynamics of the AOD process, and the estimated growth-decay merely serves as a bias correction, or compensation, to the filtered process rather than estimating the actual growth or decay of the process. 

\section{Conclusions} \label{sec:conclusions}
This paper presented a new dynamical model for spatio-temporal processes motivated from second-order SPDEs. Unlike the existing linear finite-state dynamical model for stochastic advection-diffusion process with a white-in-time Gaussian noise, the proposed dynamical model involves state variables directly related
to the process itself as well as its first-order time derivative. The proposed model was obtained by firstly constructing an infinite-dimensional SDE, followed by a finite-dimensional approximation to
the infinite-dimensional SDE using the Galerkin's method. The proposed approach has been applied to three simulation-based studies and a real case study involving the AOD propagation process of the 2025 Kenneth Fire in the Los Angeles area. The comparison study has successfully demonstrated the superior performance of the proposed dynamical model over the existing approach, making the proposed model a good candidate for process modeling and monitoring of many scientific and engineering spatio-temporal processes. 

\bibliography{sample}
\end{document}